\documentclass{article}

\usepackage{arxiv}

\usepackage[utf8]{inputenc} % allow utf-8 input
\usepackage[T1]{fontenc}    % use 8-bit T1 fonts
\usepackage{hyperref}       % hyperlinks
\usepackage{url}            % simple URL typesetting
\usepackage{booktabs}       % professional-quality tables
\usepackage{amsfonts}       % blackboard math symbols
\usepackage{nicefrac}       % compact symbols for 1/2, etc.
\usepackage{microtype}      % microtypography
\usepackage{lipsum}		% Can be removed after putting your text content
\usepackage{graphicx}
\usepackage{natbib}

\usepackage{amsmath} 
\usepackage{float}
\usepackage{subcaption}
\usepackage{threeparttablex}
\usepackage{adjustbox}
\newtheorem{prop}{Proposition}

\usepackage{booktabs}      % For \toprule, \midrule, and \bottomrule
\usepackage{siunitx}       % For formatting numbers and table alignment
\usepackage{threeparttable} % For notes under the table
\usepackage{caption}       % To control captions
\usepackage{geometry}      % For setting margins

\title{GARCH option valuation with long-run and short-run volatility components: A novel framework ensuring positive variance}

\author{ Luca Vincenzo Ballestra\\
	Department of Statistical Sciences \\
	University of Bologna \\ Bologna, Italy \\
	\And
	  Enzo D'Innocenzo \\
	Department of Economics \\
	University of Bologna \\           Bologna, Italy \\
       \And
       Christian Tezza\thanks{
   Corresponding author. Dipartimento di Scienze Statistiche, Alma Mater Studiorum Università di Bologna, Via Belle Arti 41, 40126 Bologna, Italy, e-mail: \href{mailto:christian.tezza@unibo.it}{christian.tezza@unibo.it}.}\\
  Department of Statistical Sciences \\
	University of Bologna \\ Bologna, Italy \\
}

\renewcommand{\shorttitle}{\textit{arXiv} Template}

\hypersetup{
pdftitle={A template for the arxiv style},
pdfsubject={q-bio.NC, q-bio.QM},
pdfauthor={David S.~Hippocampus, Elias D.~Striatum},
pdfkeywords={First keyword, Second keyword, More},
}

\begin{document}
\maketitle

\begin{abstract}
	\citeauthor*{christoffersen08} (\citeyear{christoffersen08}) (CJOW) proposed an improved Generalized Autoregressive Conditional Heteroskedasticity (GARCH) model for valuing European options, where the return volatility is comprised of two distinct components. Empirical studies indicate that the model developed by CJOW outperforms widely-used single-component GARCH models and provides a superior fit to options data than models that combine conditional heteroskedasticity with Poisson-normal jumps. However, a significant limitation of this model is that it allows the variance process to become negative. \cite{park23} partially addressed this issue by developing a related model, yet the positivity of the volatility components is not guaranteed, both theoretically and empirically. In this paper we introduce a new GARCH model that improves upon the models by CJOW and \cite{park23}, ensuring the positivity of the return volatility. In comparison to the two earlier GARCH approaches, our novel methodology shows comparable in-sample performance on returns data and superior performance on S\&P500 options data.
\end{abstract}

% keywords can be removed

\section{Introduction}

Building on the pioneering work of \cite{engle99}, \citeauthor*{christoffersen08} (\citeyear{christoffersen08}) (CJOW) proposed an interesting GARCH model with two volatility components, hereafter referred to as the \textit{GARCH-CJOW} model. One volatility component is a long-run component that can be modeled as a mean-reverting or as a fully persistent process, while the other is a short-run mean-reverting component with zero mean. As shown by the authors, the model allows for an efficient valuation of European options and its most distinctive feature is its ability to accurately describe the implied volatility term structure. In particular, the empirical analysis that CJOW conducted on both long-maturity and short-maturity options demonstrates superior pricing performance compared to the popular GARCH model with a single volatility component developed by \cite{heston00} and a GARCH(1,1) model augmented with Poisson-normal jumps. Owing to its well-documented empirical performance, the \textit{GARCH-CJOW} model has gained popularity in the GARCH literature and has been used for pricing S\&P500 options in \cite{corsi13} and VIX futures in \cite{cheng23}.

Despite its success, \cite{bormetti15} and \cite{park23} provide evidence that the \textit{GARCH-CJOW} model fails to guarantee positive volatilities for parameter sets commonly encountered in financial practice. This raises significant modeling concerns for out-of-sample (volatility) analysis and for option pricing, especially for medium to long-maturing options. Even though \cite{bormetti15} state that the likelihood of obtaining negative volatilities is extremely low, \cite{park23} further analyze the performances of the \textit{GARCH-CJOW} model and find that it can generate a consistent number of negative volatility trajectories. Then, to address the issue, they propose an enhancement of the \textit{GARCH-CJOW} model, hereafter referred to as \textit{GARCH-OP}, which allows for the pricing of SPX options and VIX derivatives. However, this model also does not ensure positivite volatility.

In this paper, we investigate further the issue of negative volatility of the \textit{GARCH-CJOW} model and its implications for option pricing. Contrary to the conclusions drawn by \cite{bormetti15}, we demonstrate that, for parameter sets established in the empirical literature, the \textit{GARCH-CJOW} model generates a consistent number of negative volatility trajectories. Additionally, the \textit{GARCH-OP} model also produces negative volatilities, a concern that becomes more pronounced for longer time horizons. We also find that this issue affects the computation of option prices, as it is based on an inversion formula where the integrands for longer maturities may diverge. %Consequently, the conventional semi-closed formula for calculating option prices is not applicable, requiring the use of Monte Carlo (MC) simulations with modifications to prevent negative volatilities. Specifically, when conducting MC simulations, one may enforce positive volatility by introducing an absolute value into the model equations. However, this adjustment alters the model equations, potentially leading to different and possibly undesirable statistical properties.

To fix these issues, we propose an improvement to the model developed by CJOW that ensures the positivity of the volatility process by properly specifying the impact of the innovation on volatility. This new model, which allows for a well-posed option semi-closed form valuation, will be referred to as corrected positive component GARCH model, or, in short, \textit{GARCH-CPC}. In contrast to the \textit{GARCH-CJOW} model, the semi-closed formula of the \textit{GARCH-CPC} model can be safely used in every option pricing scenario, even for large option maturities. 

Moreover, we compare the performance of the \textit{GARCH-CJOW}, \textit{GARCH-CPC} and \textit{GARCH-OP} models in the valuation of a large panel of options written on the S\&P500 index. We find that the novel \textit{GARCH-CPC} model offers a feasible and efficient solution for computing option prices using affine formulas with greater accuracy than both the \textit{GARCH-CJOW} and the \textit{GARCH-OP} models.

The remainder of this paper is structured as follows. In Section \ref{sec_model}, we briefly recall the \textit{GARCH-CJOW} and the \textit{GARCH-OP} models, showing the negativity issues affecting these approaches and their implications for option pricing. Section \ref{sec_positive} proposes the \textit{GARCH-CPC} model. After deriving the risk-neutral dynamics, we provide the formulas for option valuation.
Section \ref{sec_empirical} presents an empirical study conducted on returns and option data. Finally, Section \ref{sec_conclusions} concludes.

\section{Return dynamics with volatility components} \label{sec_model}

The volatility component model proposed by CJOW for the return process $R_t = \ln{\left(\frac{S_t}{S_{t-1}}\right)}$, where $S_t$ denotes the spot price, is given by the following equations (under the physical measure): 

\begin{align}  
    \label{eq_cjow_rt}
    R_{t+1} &= r + \lambda h_{t+1} + \sqrt{h_{t+1}}  Z_{t+1}, \\ \label{eq_cjow_ht}
    h_{t+1} &= q_{t+1} + \tilde{\beta} (h_{t}-q_{t}) + \alpha \left(Z^2_{t}-1-2\gamma_1 \sqrt{h_{t}}Z_{t} \right), \\ 
    \label{eq_cjow_qt}
    q_{t+1} &= \omega + \rho q_{t} + \varphi \left(Z^2_{t}-1-2\gamma_2 \sqrt{h_{t}}Z_{t} \right), 
\end{align}

\noindent where $r$ denotes the risk-free rate, $Z_t \overset{iid}{\sim}N(0,1)$. CJOW refer to $q_t$ and $h_t - q_t$ as the long-run and the short-run variance component, respectively. They impose the parameter constraints $\omega \geq 0$, $\alpha>0$, $\varphi>0$, $\tilde{\beta} < 1$ and  $\rho  < 1$. In the long-run we have that $\mathbb{E}[h_t] = \mathbb{E}[q_t] = {\omega}/(1-\rho)$.

As pointed out in \cite{bormetti15} and \cite{park23}, the model \eqref{eq_cjow_rt}-\eqref{eq_cjow_qt} does not guarantee that the total variance $h_t$ remains positive. Therefore, in \cite{park23}, the following model, hereafter referred to as \textit{GARCH-OP}, has been proposed as a more robust alternative to address the negative volatility issue that affects the \textit{GARCH-CJOW} model: %To provide a direct and practical understanding of this fact, let us consider the case that $\omega<\frac{\varphi}{2}$, as is typically estimated based on market data (see CJOW and \cite{cheng23}), and consider the trajectory such that $Z_t = 0$ for all $t$. From equation \eqref{eq_cjow_qt} we deduce that, for sufficiently large $t$, $q_t$ becomes negative, and, consequently, also $h_t$ in equation \eqref{eq_cjow_ht} becomes negative. 

\begin{align}  
    \label{eq_park_rt}
    R_{t+1} &= r + \lambda h_{t+1} + \sqrt{h_{t+1}}  Z_{t+1}, \\ \label{eq_park_ht}
    h_{t+1} &= q_{t+1} + \tilde{\beta} (h_{t}-q_{t}) + \alpha \left(Z_{t}-\gamma_1 \sqrt{h_{t}} \right)^2 - \omega - \alpha \gamma^2_1 h_t, \\ 
    \label{eq_park_qt}
    q_{t+1} &= \omega + \rho q_{t} + \varphi \left(Z_{t}-\gamma_2 \sqrt{h_{t}} \right)^2, 
\end{align}

\noindent where the long-term means are given by $\begin{pmatrix}
        \mathbb{E}[h_t] &
        \mathbb{E}[q_t]
    \end{pmatrix}^{\top} =  (I_2 - P)^{-1} R$, where
    $I_2$ denote a $2\times2$ identity matrix, $P= \begin{pmatrix}
     \Tilde{\beta}  + \varphi \gamma^2_2& \rho - \Tilde{\beta}  \\
     \varphi \gamma^2_2 & \rho 
 \end{pmatrix}$ and $R= \begin{pmatrix}
     \alpha + \varphi \\ 
      \omega + \varphi
 \end{pmatrix}$.

\subsection{Model issues}

In this section, we will document the issues of model \eqref{eq_cjow_rt}-\eqref{eq_cjow_qt} and model \eqref{eq_park_rt}-\eqref{eq_park_qt} related to the negative values that the variance can take through a few test cases. Using parameters established in the empirical literature, we will show via statistical simulation how many variance trajectories $h_t$ can turn negative, and we will examine the implications for option pricing.

\subsubsection{Negative volatility trajectories}

For the model \eqref{eq_cjow_rt}-\eqref{eq_cjow_qt}  we initially consider two sets of parameters available in the literature, namely, those in \cite{christoffersen08} and in \cite{cheng23}, which we denote as \textit{CJOW08} and \textit{CCLT23}, respectively. These parameters are reported in Table \ref{tab:param}. We note that equation (1) in \cite{cheng23} specifies $\lambda-\frac{1}{2}$ instead of just $\lambda$ as in equation \eqref{eq_cjow_rt} in the present paper, so in Table \ref{tab:param} we adjusted the value of $\lambda$ accordingly. Whereas, for model in \eqref{eq_park_rt}-\eqref{eq_park_qt} we consider the same parameters as in \cite{park23}, which we denote \textit{OP23} and we report in Table \ref{tab:param}.

\begin{table}[h]
    \centering
       \caption{Model parameters.}
    \label{tab:param}
    \resizebox{\textwidth}{!}{
    \begin{tabular}{c c c c c c c c c}
    \toprule
      {Parameter} & {$\omega$} & {$\alpha$} & {$\gamma_1$} & {$\tilde{\beta}$} & {$\varphi$} & {$\gamma_2$} & {$\rho$} & {$\lambda$} \\
       \midrule
       { \textit{CJOW08}} & 8.208{e-07} & 1.580{e-06} & 415.100 & 0.6437 & 2.480{e-06} & 63.240 & 0.9896 & 2.092 \\
       {\textit{CCLT23}} & 7.776e-07 & 1.380e-06 & 402.352 & 0.862 & 1.795e-06 & 73.205 & 0.991 & 1.357 \\
       {\textit{OP23}} & -1.57e-06 & 0.190{e-06} & 7050 & 0.922 & 2.62e-06 & 89 & 0.983 & -7.88 \\
    \bottomrule
    \end{tabular}
    }
\end{table}

We perform a Monte Carlo exercise where we simulate the trajectories of the variance in equations \eqref{eq_cjow_ht} and \eqref{eq_park_ht} to check if, and how many of them, turn negative. The simulation consists in generating 1,000,000 paths of innovations ${Z_t}$ from $t=0$ to a given time horizon $T>0$. Using equations \eqref{eq_cjow_ht}-\eqref{eq_cjow_qt} and \eqref{eq_park_ht}-\eqref{eq_park_qt}, for each path we compute the variance $h_t$ and count the number of trajectories such that $h_t$ becomes negative for some $t \leq T$.  We set $S_0 = 100$, $r = 10^{-5}$ and, following \cite{park23}, we choose $q_0$ and $h_0$ such that the (annualized) initial volatility states $\sqrt{252 q_0}$ and $\sqrt{252 h_0}$ are equal to $5\%$ and $10\%$, respectively. The results, illustrated in Table \ref{tab:negative_vols}, show that as $T$ increases, the number of negative trajectories also increases for both the \textit{GARCH-CJOW} and \textit{GARCH-OP} models. However, as documented in \cite{park23}, the number of negative trajectories generated by the \textit{GARCH-OP} model is significantly lower compared to the \textit{GARCH-CJOW} model.

\begin{table}[h]
    \centering
    \caption{Negative trajectories for the variance $h_t$ out of 1,000,000 simulations for time horizon $T$.}
    \label{tab:negative_vols}
    \begin{tabular}{c c c c c c c}
        \toprule
        & \multicolumn{6}{c}{Initial volatility
        $\left(\sqrt{252 q_0} , \sqrt{252 h_0}\right)$} \\  
        \cmidrule(lr){2-7} 
        & \multicolumn{3}{c}{(5\% , 5\%)}  & \multicolumn{3}{c}{(10\% , 10\%)} \\
         \cmidrule(lr){2-4} \cmidrule(lr){5-7}
        & \multicolumn{2}{c}{\textit{GARCH-CJOW}} & \multicolumn{1}{c}{\textit{GARCH-OP}}  & \multicolumn{2}{c}{\textit{GARCH-CJOW}} & \multicolumn{1}{c}{\textit{GARCH-OP}} \\
        \cmidrule(lr){2-3} \cmidrule(lr){4-4} \cmidrule(lr){5-6} \cmidrule(lr){7-7}
         {$T$} &   {\textit{CJOW08} }  & {\textit{CCLT23} } & {\textit{OP23} }  & {\textit{CJOW08} }     & {\textit{CCLT23} }   & {\textit{OP23} } \\
         \midrule
         {15} & 226,386 & 185,403 & 2,328 & 0 & 0 & 0 \\
         {30} & 287,888 & 235,937 & 7,848 & 317 & 0 & 0 \\
         {50} & 315,161 & 251,841 & 10,422 & 3,671 & 115 & 7 \\
         {80} & 330,745 & 258,352 & 11,023 & 10,034 & 481 & 33 \\
         {120} & 339,795 & 260,234 & 11,112 & 16,883 & 891 & 41 \\
         {252} & 351,374 & 261,183 & 11,114 & 29,129 & 1,465 & 44 \\
         \bottomrule
    \end{tabular}    
\end{table}

\subsubsection{Option pricing}\label{sec_opt_pricing}

In this subsection, after briefly recalling formulas commonly used to compute option prices, we empirically analyze the consequences of generating negative volatility trajectories for option valuation in the \textit{GARCH-CJOW} and \textit{GARCH-OP} models.
 
As shown by \cite{heston00} and further utilized by CJOW, the inversion formula developed in \cite{gilpelaez51} yields the following expression for a European call option on a non-dividend paying stock with spot price $S_t$, strike price $K$ and expiration date $T$:

\begin{align} \label{eq_opt_price}
    C_t = \frac{1}{2} S_t &+ \frac{e^{-r(T-t)}}{\pi} \int_{0}^{\infty} \text{Re} \left[ \frac{ K^{-i \phi} f^*(t,T; i \phi + 1) }{ i \phi} \right] d\phi \\ \notag
    &- K e^{-r(T-t)} \left(\frac{1}{2} + \frac{1}{\pi}
    \int_{0}^{\infty} \text{Re} \left[ \frac{ K^{-i \phi} f^*(t,T; i \phi ) }{ i \phi } \right] d\phi  \right),
\end{align}

\noindent where $\text{Re}[\cdot]$ denotes the real part of a complex number and $f^*(t,T; i \phi )$ represents the conditional characteristic function of the terminal log-stock price under the risk-neutral measure. The risk-neutral conditional characteristic function of the \textit{GARCH-CJOW} and \textit{GARCH-OP} models can be computed as follows: 

\begin{equation} \label{eq:cf}
    f^*(t,T;\phi) = S_t^\phi \exp{ \{A_t(\phi) + B_{1,t}(\phi) (h_{t+1} - q_{t+1}) + B_{2,t}(\phi) q_{t+1}\}}, \quad t < T,
\end{equation}

\noindent where the expressions of $A_t(\phi)$, $B_{1,t}(\phi)$ and $B_{2,t}(\phi)$ can be obtained using equations (25) in the paper by \cite{christoffersen08} for the \textit{GARCH-CJOW} model and using equations (8) in the paper by \cite{park23} for the \textit{GARCH-OP} model. 

Let us now focus on the integrand functions in \eqref{eq_opt_price}:

\begin{align} \label{eq_integrands}
    I_1 =  \text{Re} \left[ \frac{ K^{-i \phi} f^*(t,T; i \phi + 1) }{ i \phi} \right], \quad 
     I_2 = \text{Re} \left[ \frac{ K^{-i \phi} f^*(t,T; i \phi ) }{ i \phi} \right].
\end{align}

For both the \textit{GARCH-CJOW} and \textit{GARCH-OP} models, we check the integrability of $I_1$ and $I_2$, by analyzing the behavior of $I_1$ and $I_2$ as functions of $\phi$ and for different option maturities $T=15,30,50,80,120,252$. For comparison purposes, we also consider the values of $I_1$ and $I_2$ for the popular GARCH model developed in \cite{heston00}, hereafter \textit{GARCH-HN}:

\begin{align}
    \label{eq_hn_rt}
    R_{t+1} &= r + \lambda h_{t+1} + \sqrt{h_{t+1}}  Z_{t+1}, \\ \label{eq_hn_ht}
    h_{t+1} &= \omega + \tilde{\beta} h_{t} + \alpha \left(Z_{t}-\gamma_1 \sqrt{h_{t}} \right)^2.  
\end{align}

Similarly to the \textit{GARCH-CJOW} model, for the \textit{GARCH-HN} model we use parameters obtained by \cite{christoffersen08} and \cite{cheng23}, which are reported in Table \ref{tab:param2}. We note that, for the \textit{GARCH-HN} model, option prices can be computed using equation \eqref{eq_opt_price} in conjunction with the risk-neutral conditional characteristic function obtained by \cite{heston00}.

\begin{table}[h]
    \centering
       \caption{Parameters of the \textit{GARCH-HN} obtained by CJOW and \cite{cheng23}.}
    \label{tab:param2}
    \begin{tabular}{c c c c c c}
    \toprule
      {Parameter} & {$\omega$} & {$\alpha$} & {$\gamma_1$} & {$\tilde{\beta}$} & {$\lambda$} \\
       \midrule
       { \textit{CJOW08}} & 2.101e-17 & 3.317e-06 & 1.276e+02 & 0.9552 & 2.231 \\
       { \textit{CCLT23}} & 1.744e-06 & 3.098e-06 & 120.967 & 0.935 & 1.395 \\
    \bottomrule
    \end{tabular}
\end{table}

\begin{figure}[H]
    \caption{Behavior of $I_1$ (left) and $I_2$ (right), as functions of $\phi$ for different values of $T$ for the \textit{GARCH-HN} model. We set $S_0 = K = 100$, $r = 10^{-5}$ and $\sqrt{252 h_0} = 5\%$. We use the \textit{CJOW08} parameters (top panels) and the \textit{CCLT23} parameters as in Table \ref{tab:param} (bottom panels).}
    \label{fig:diverg_hn_05}
    \centering

    \begin{subfigure}{\textwidth}
        \centering
        \includegraphics[width=0.49\textwidth]{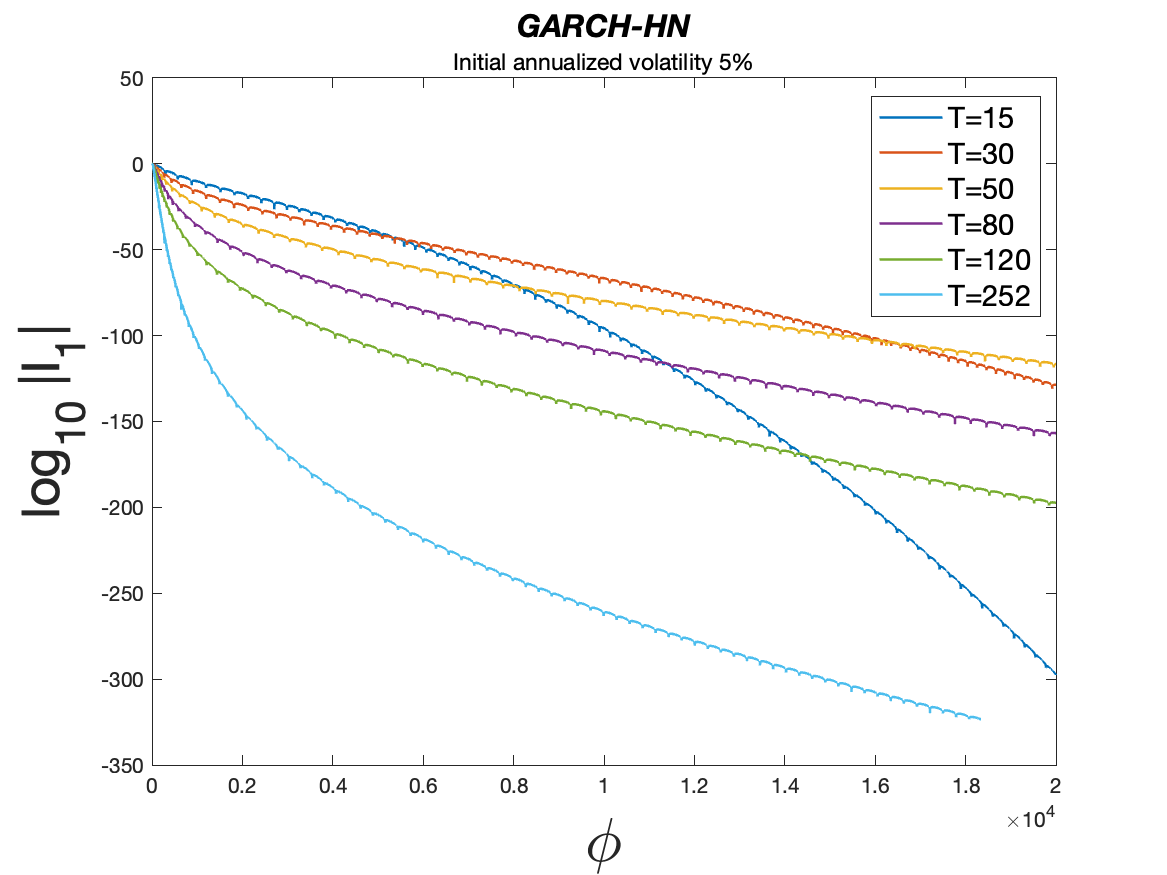}\hfill
        \includegraphics[width=0.49\textwidth]{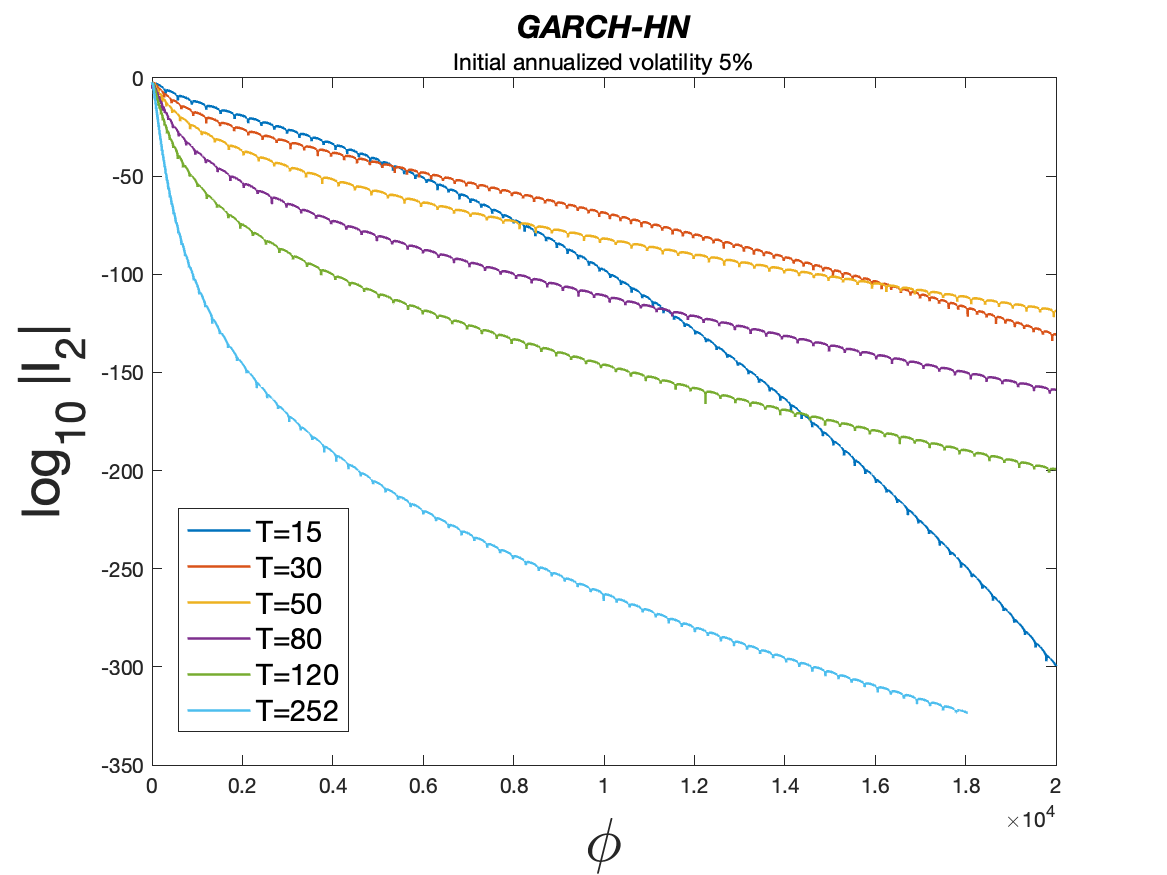}
    \end{subfigure}

    \vspace{0.5cm}

    \begin{subfigure}{\textwidth}
        \centering
        \includegraphics[width=0.49\textwidth]{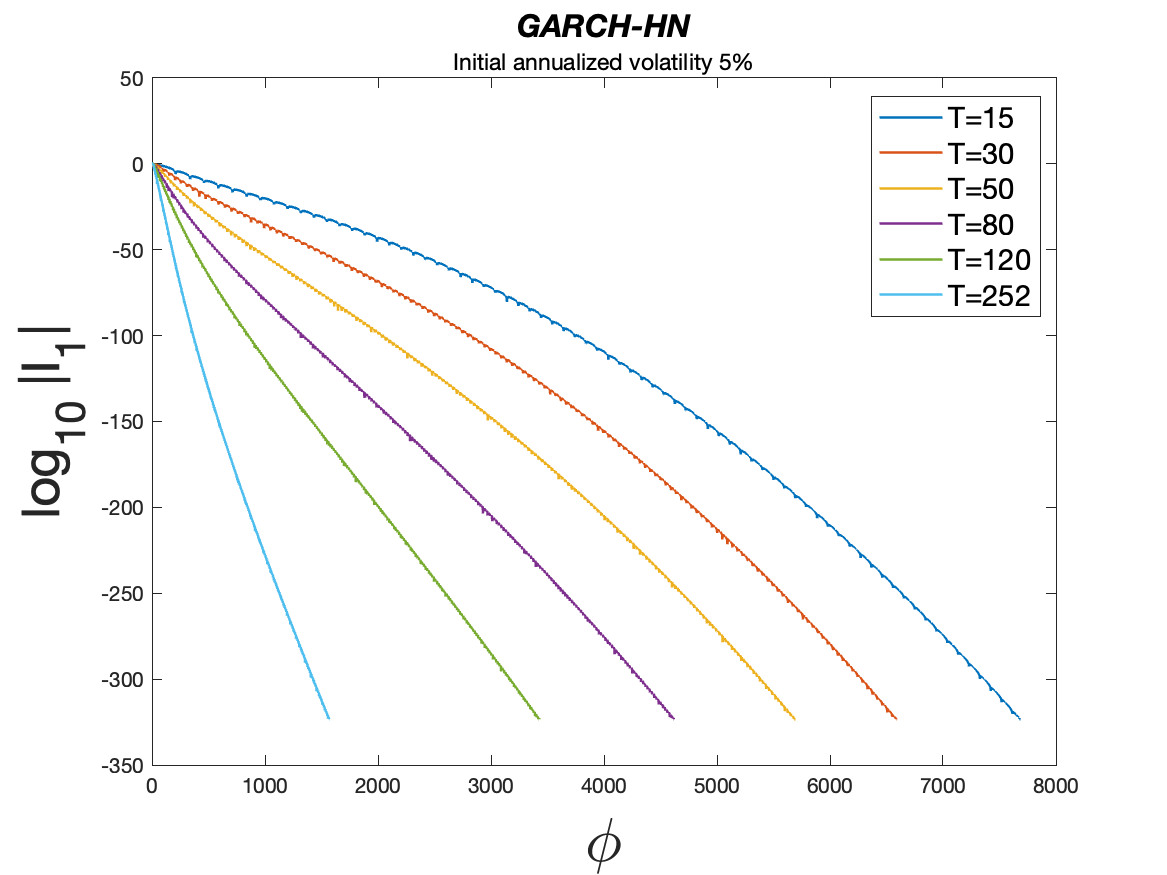}\hfill
        \includegraphics[width=0.49\textwidth]{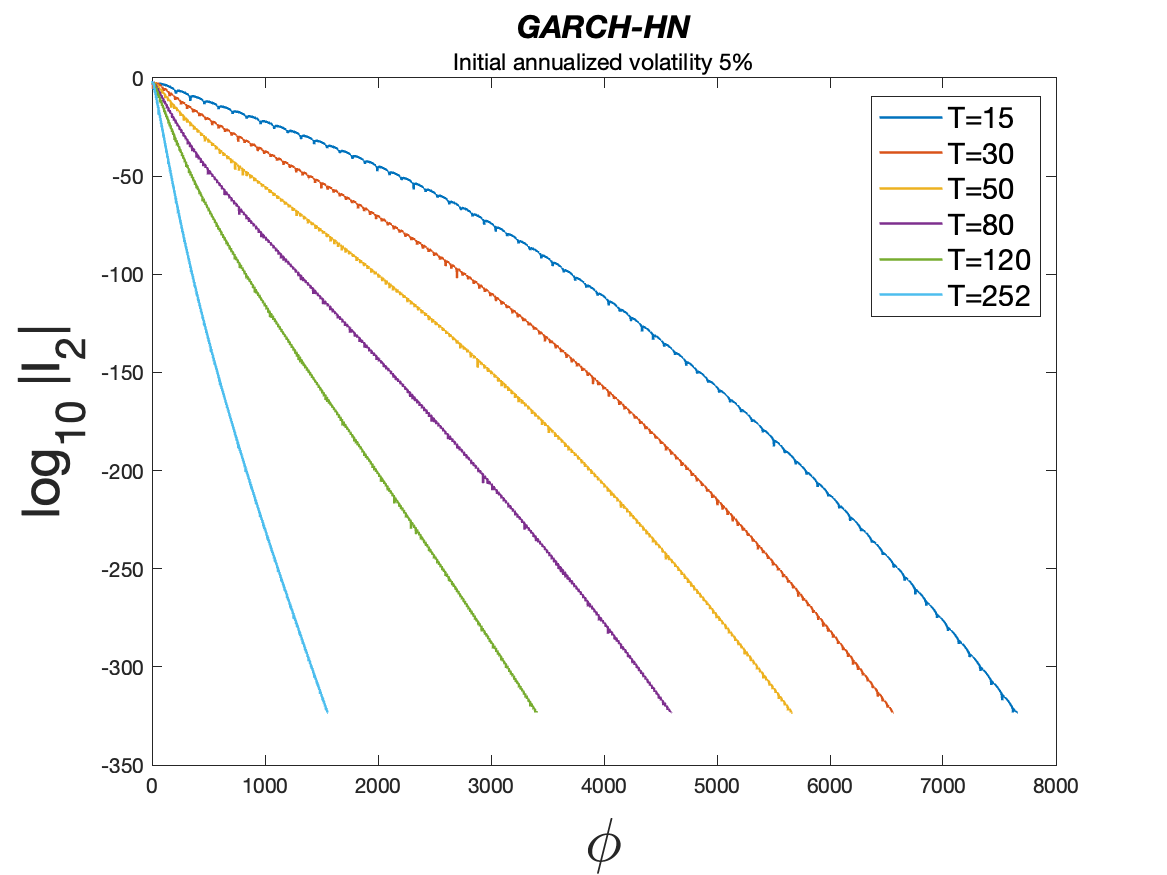}
        %\label{fig:plot3}
    \end{subfigure}
\end{figure}

\begin{figure}[H]
    \caption{Behavior of $I_1$ (left) and $I_2$ (right), as functions of $\phi$ for different values of $T$ for the \textit{GARCH-HN} model. We set $S_0 = K = 100$, $r = 10^{-5}$ and $\sqrt{252 h_0} = 10\%$. We use the \textit{CJOW08} parameters (top panels) and the \textit{CCLT23} parameters as in Table \ref{tab:param} (bottom panels).}
    \label{fig:diverg_hn_10}
    \centering

    \begin{subfigure}{\textwidth}
        \centering
        \includegraphics[width=0.49\textwidth]{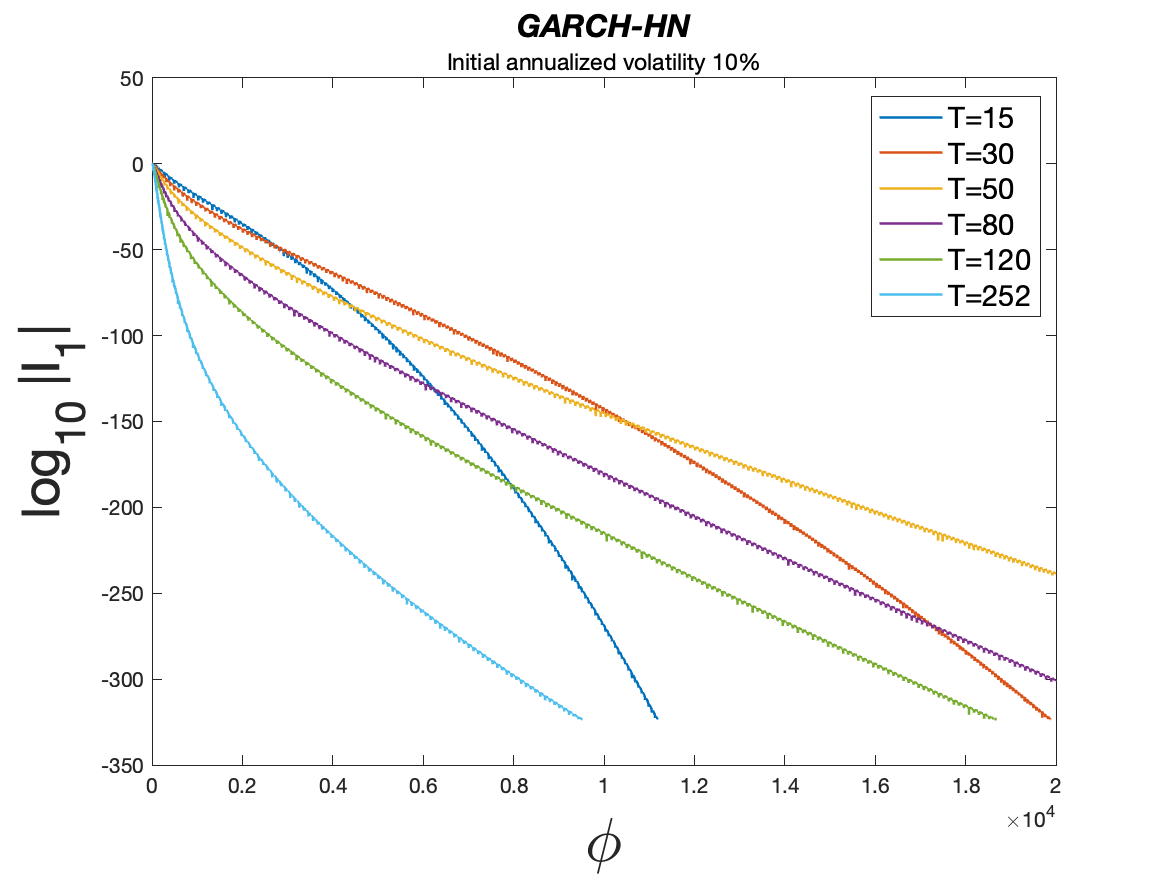}\hfill
        \includegraphics[width=0.49\textwidth]{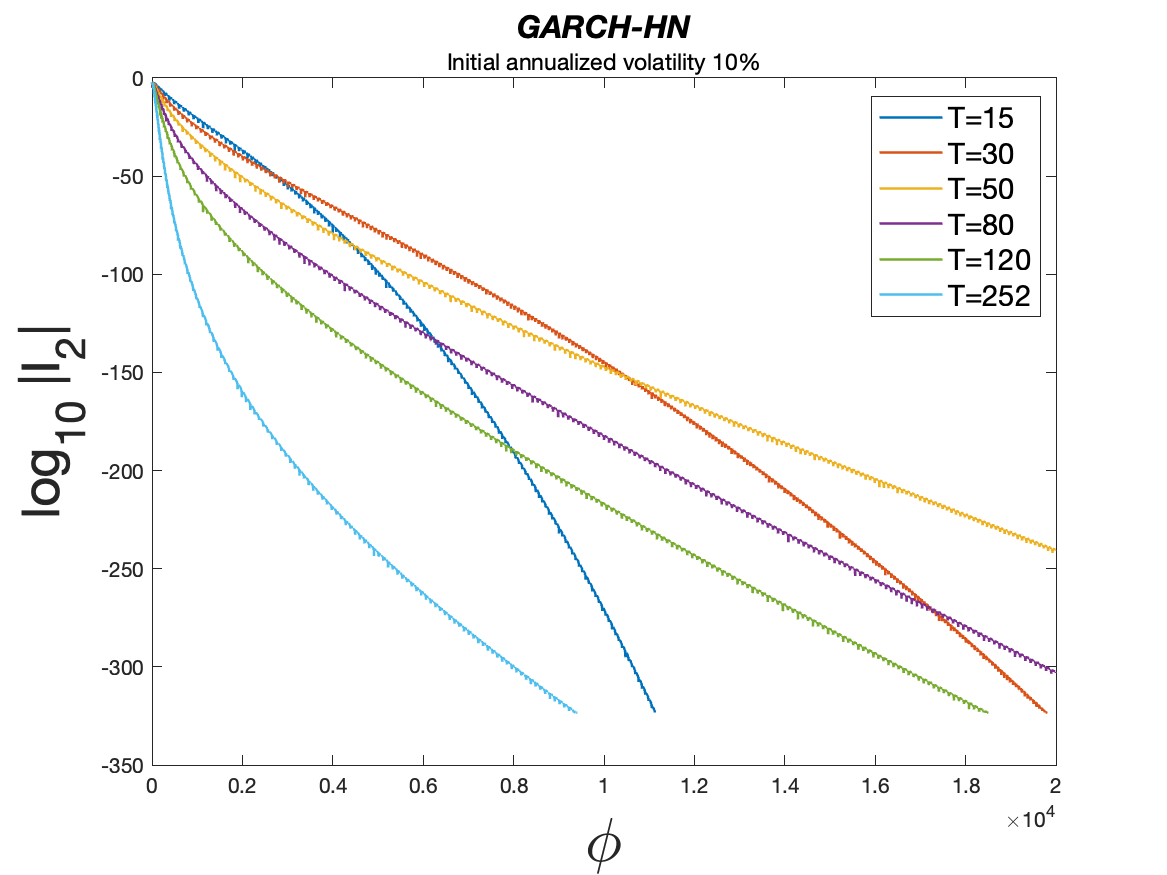}
        %\label{fig:plot1}
    \end{subfigure}

    \vspace{0.5cm}

    \begin{subfigure}{\textwidth}
        \centering
        \includegraphics[width=0.49\textwidth]{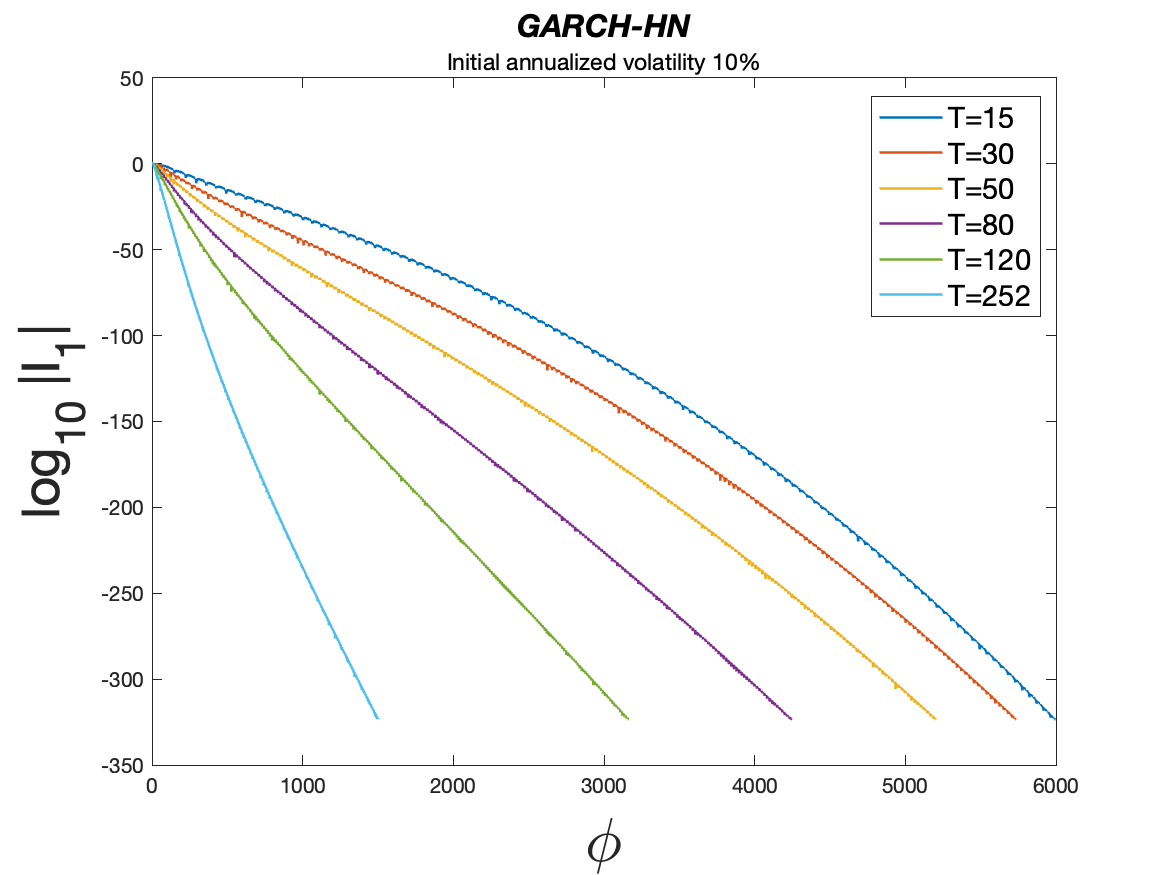}\hfill
        \includegraphics[width=0.49\textwidth]{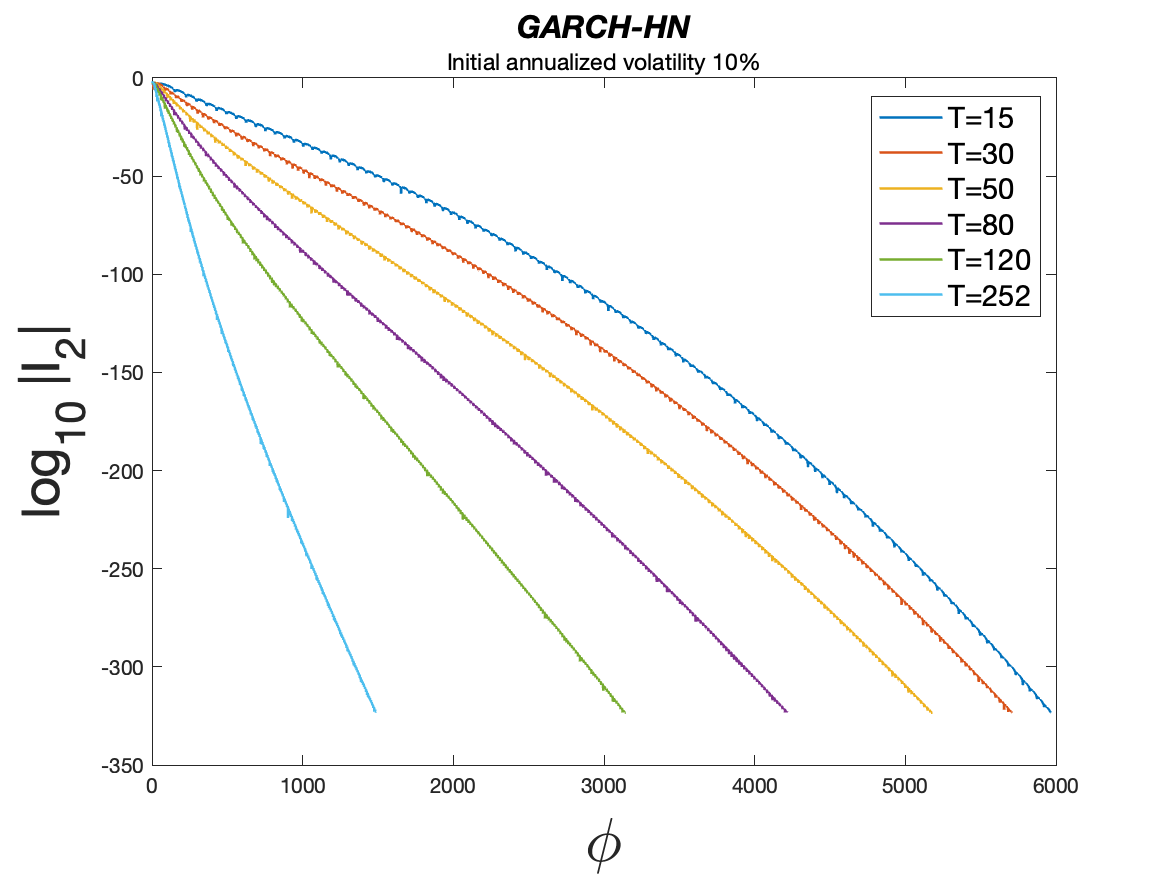}
        %\label{fig:plot3}
    \end{subfigure}
\end{figure}

\begin{figure}[H]
    \caption{Behavior of $I_1$ (left) and $I_2$ (right), as functions of $\phi$ for different values of $T$ for the \textit{GARCH-CJOW} model. We used $S_0 = K = 100$, $r = 10^{-5}$ and the initial volatility states $\sqrt{252 h_0} = 5\%$ and $\sqrt{252 q_0} = 5\%$. We use the \textit{CJOW08} parameters (top panels) and the \textit{CCLT23} parameters as in Table \ref{tab:param} (bottom panels).}
    \label{fig:diverg_cjow_05}
    \centering

    \begin{subfigure}{\textwidth}
        \centering
        \includegraphics[width=0.49\textwidth]{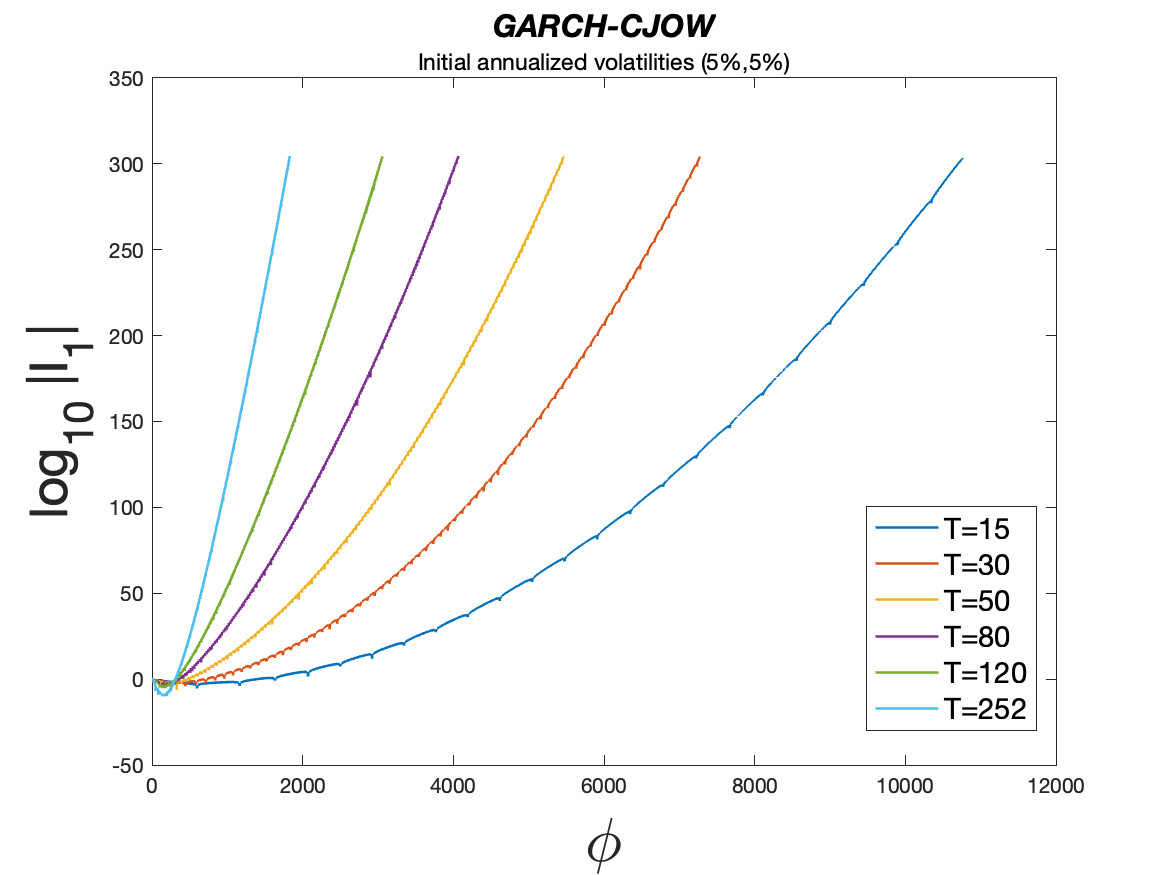}\hfill
        \includegraphics[width=0.49\textwidth]{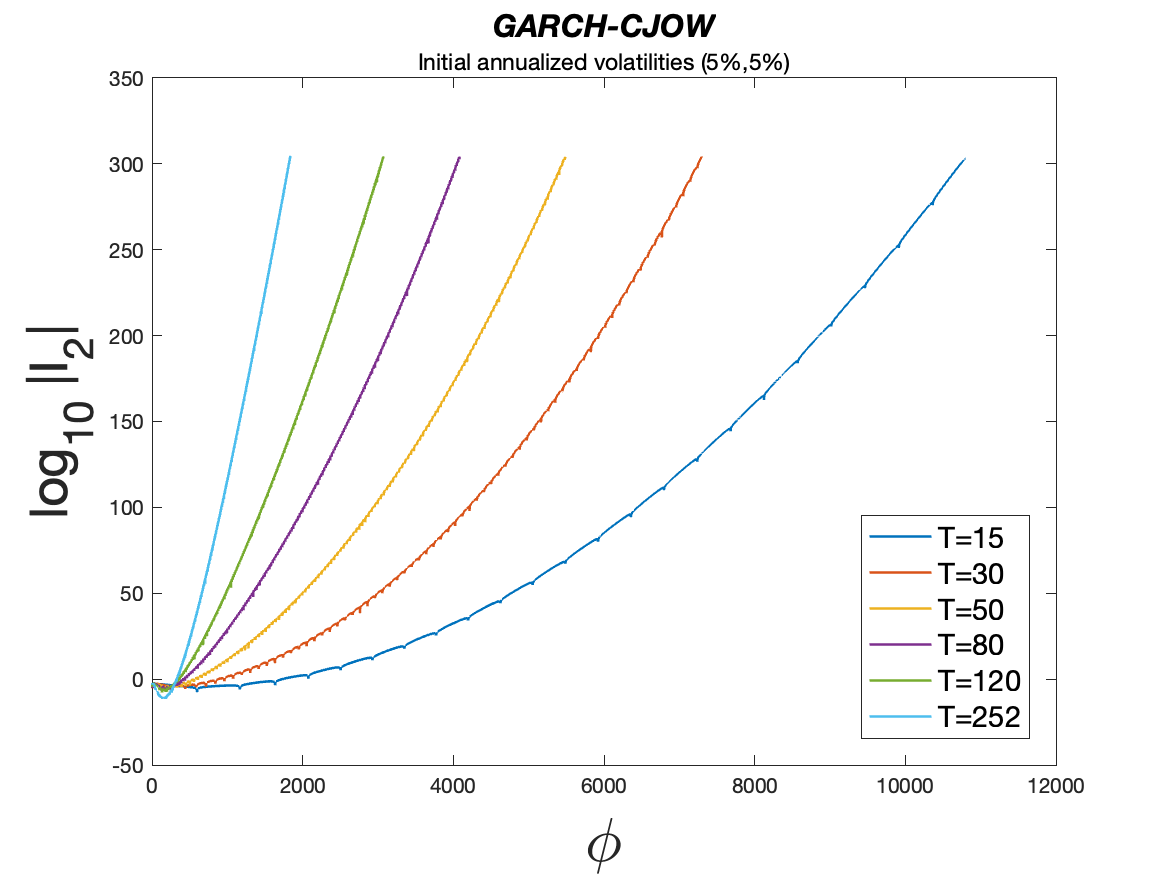}
        %\label{fig:plot1}
    \end{subfigure}

    \vspace{0.5cm}

    \begin{subfigure}{\textwidth}
        \centering
        \includegraphics[width=0.49\textwidth]{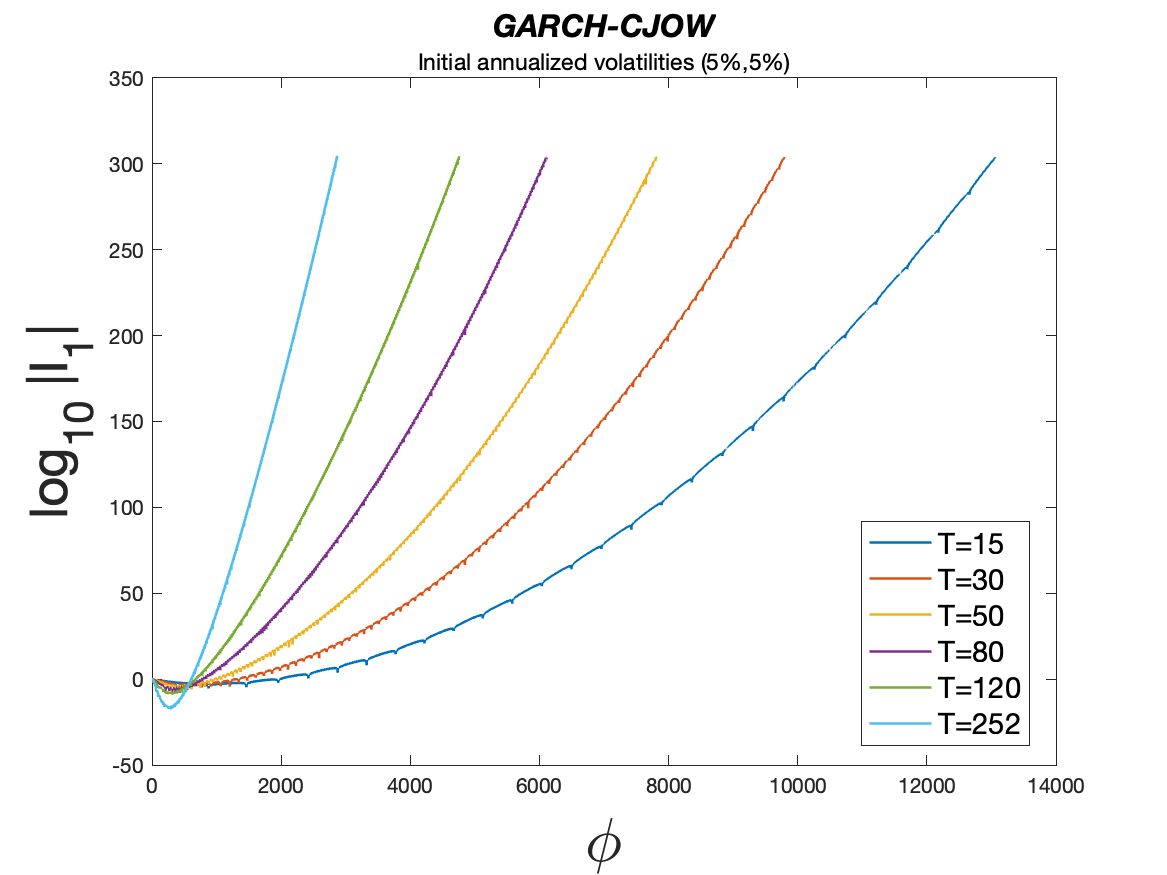}\hfill
        \includegraphics[width=0.49\textwidth]{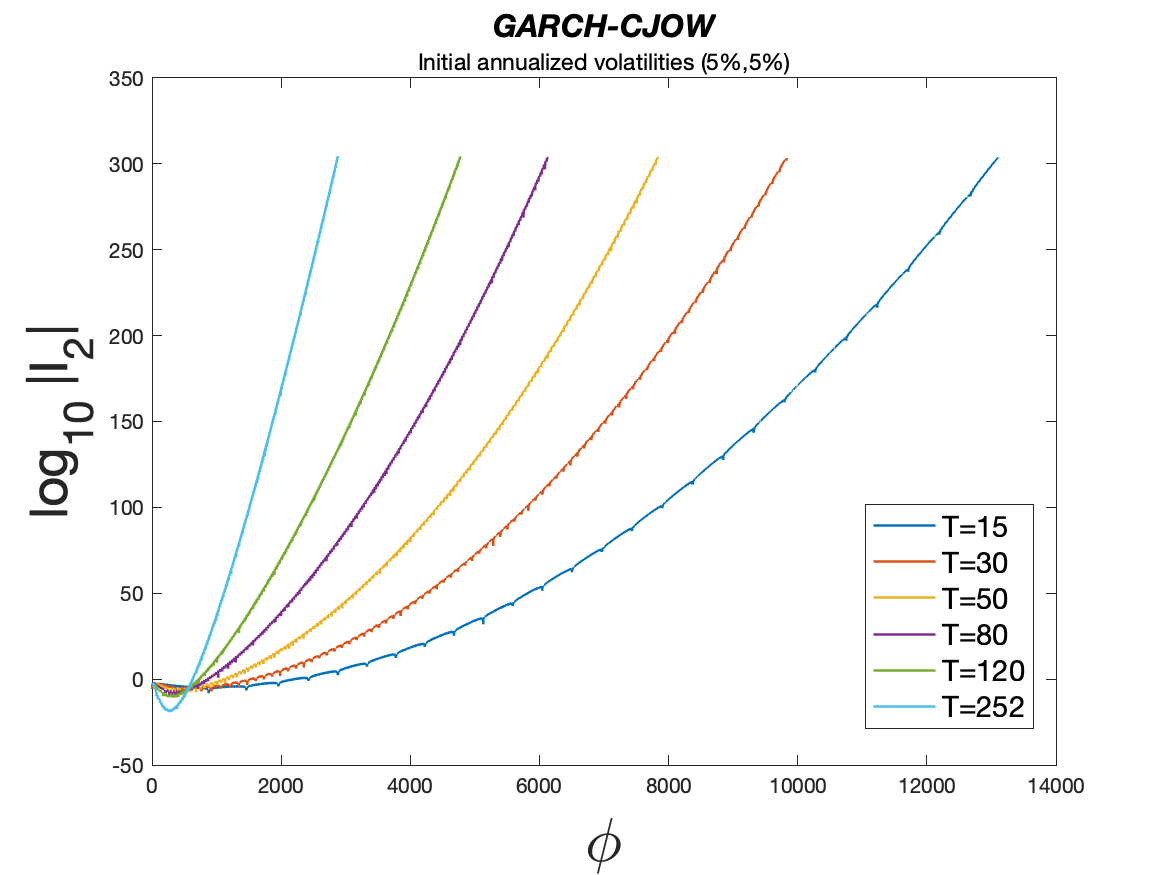}
        %\label{fig:plot3}
    \end{subfigure}
\end{figure}

\begin{figure}[H]
    \caption{Behavior of $I_1$ (left) and $I_2$ (right), as functions of $\phi$ for different values of $T$ for the \textit{GARCH-CJOW} model. We used $S_0 = K = 100$, $r = 10^{-5}$ and the initial volatility states $\sqrt{252 h_0} = 10\%$ and $\sqrt{252 q_0} = 10\%$. We use the \textit{CJOW08} parameters (top panels) and the \textit{CCLT23} parameters as in Table \ref{tab:param} (bottom panels).}
    \label{fig:diverg_cjow_10}
    \centering

    \begin{subfigure}{\textwidth}
        \centering
        \includegraphics[width=0.49\textwidth]{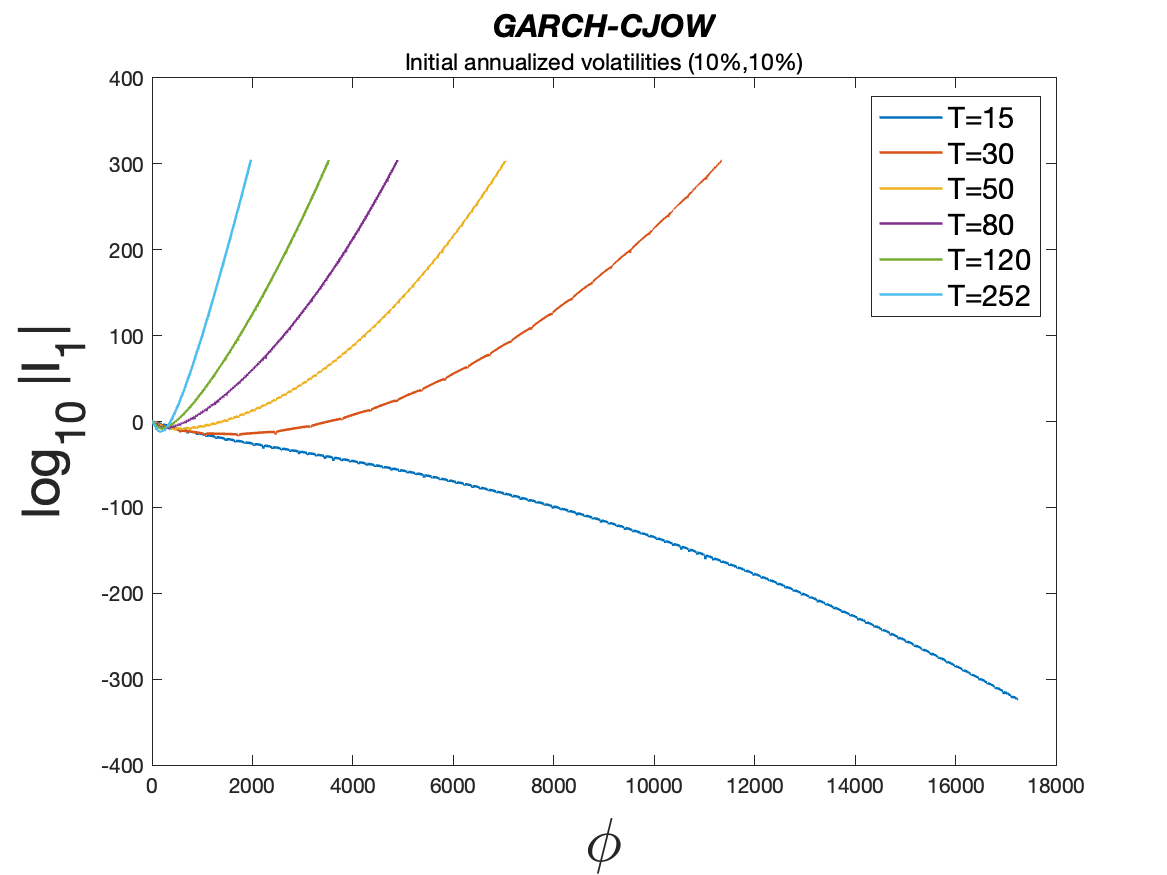}\hfill
        \includegraphics[width=0.49\textwidth]{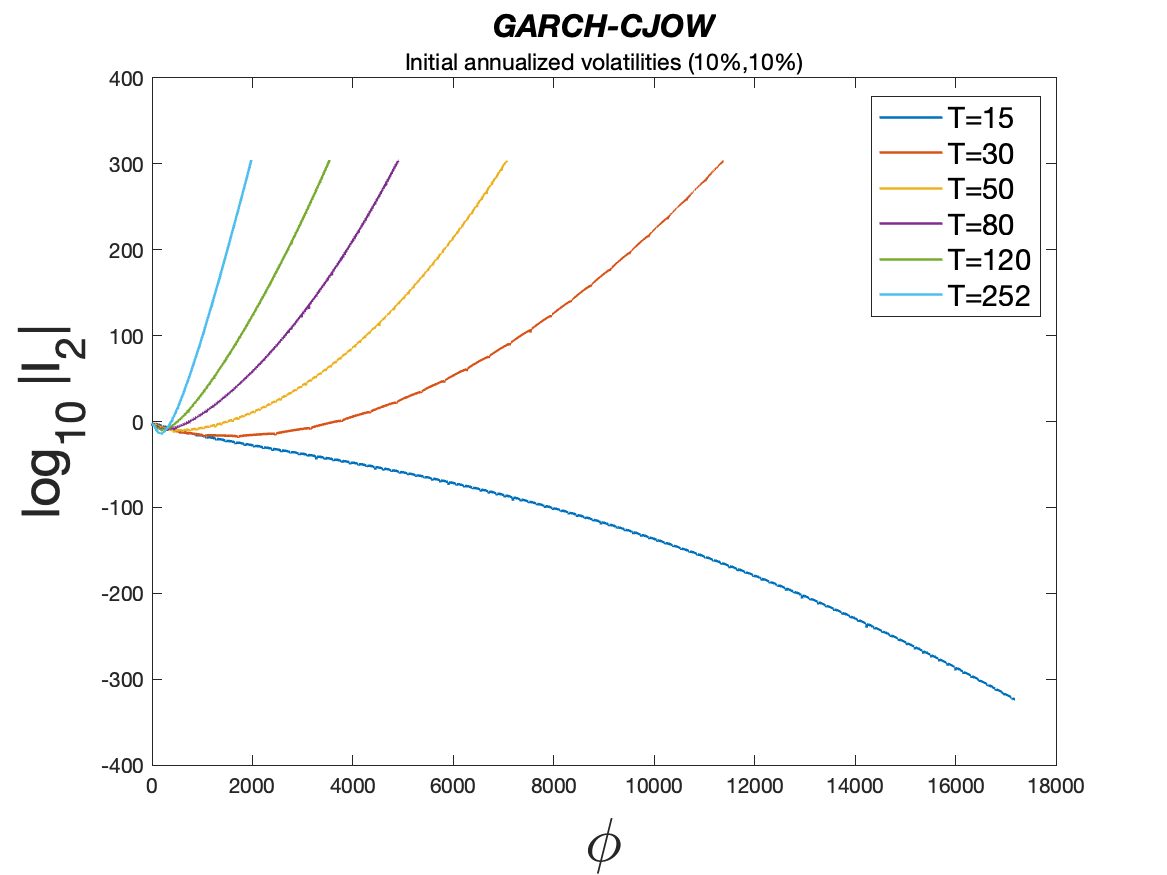}
        %\label{fig:plot1}
    \end{subfigure}

    \vspace{0.5cm}

    \begin{subfigure}{\textwidth}
        \centering
        \includegraphics[width=0.49\textwidth]{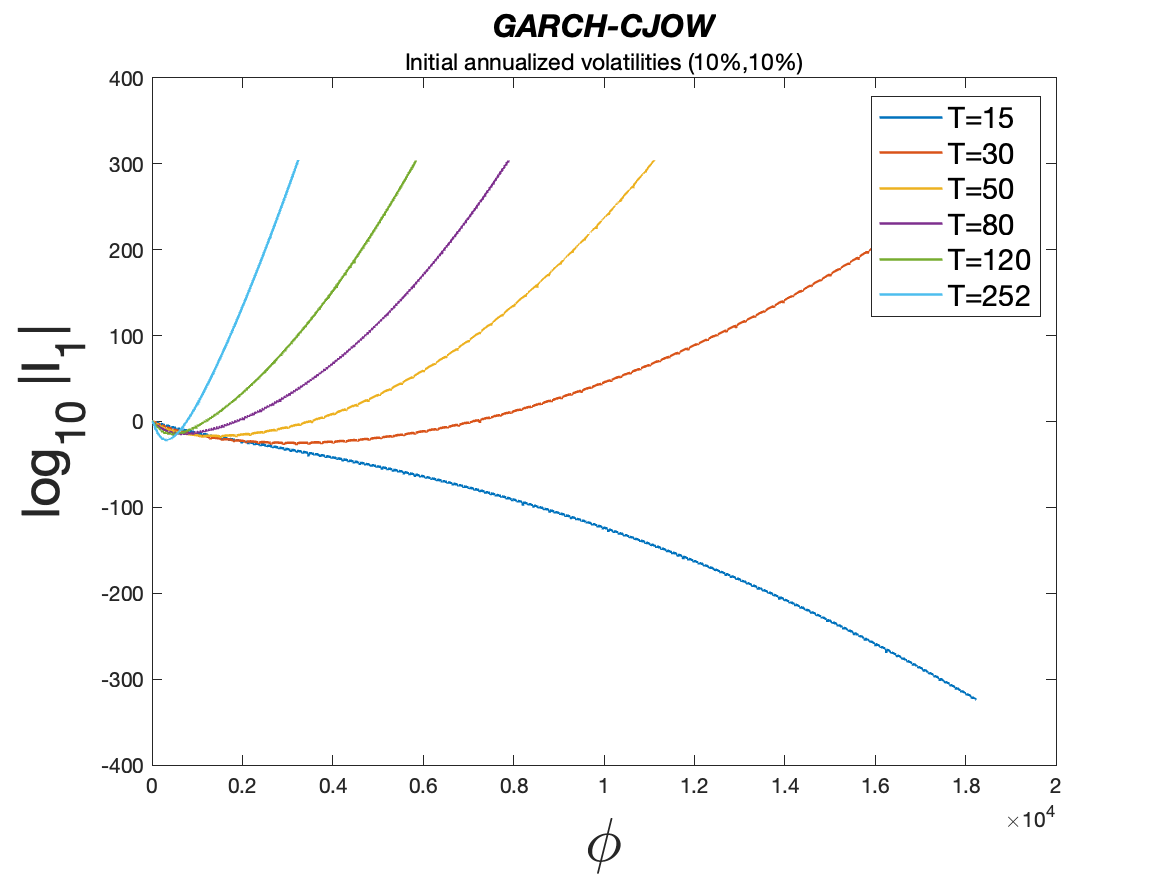}\hfill
        \includegraphics[width=0.49\textwidth]{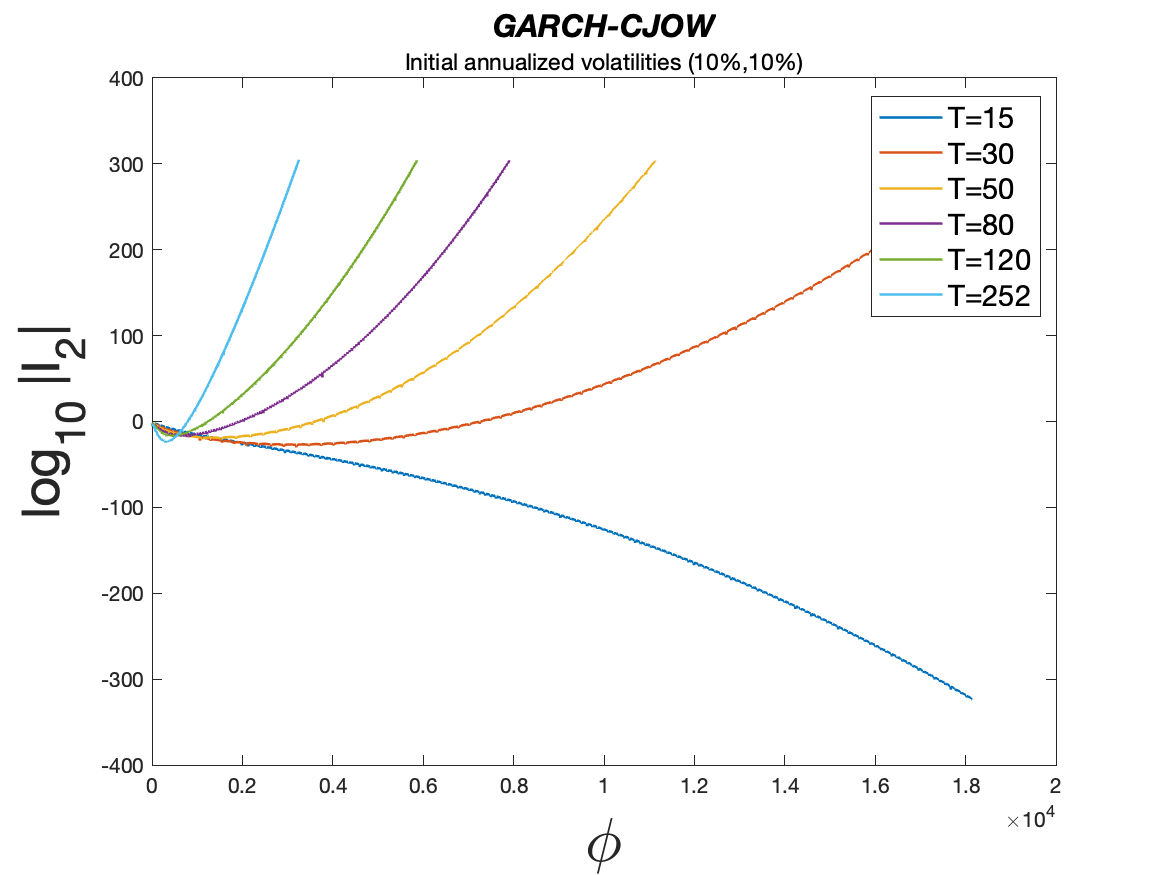}
        %\label{fig:plot3}
    \end{subfigure}
\end{figure}

\begin{figure}[H]
    \caption{Behavior of $I_1$ (left) and $I_2$ (right), as functions of $\phi$ for different values of $T$ for the \textit{GARCH-OP} model with the parameters \textit{OP23} of Table \ref{tab:param}. We set $S_0 = K = 100$, $r = 10^{-5}$. We use $\sqrt{252 h_0} = \sqrt{252 q_0} = 5\%$ (top panels) and $\sqrt{252 h_0} = \sqrt{252 q_0} = 10\%$ (bottom panels).}
    \label{fig:diverg_park}
    \centering

    \begin{subfigure}{\textwidth}
        \centering
        \includegraphics[width=0.49\textwidth]{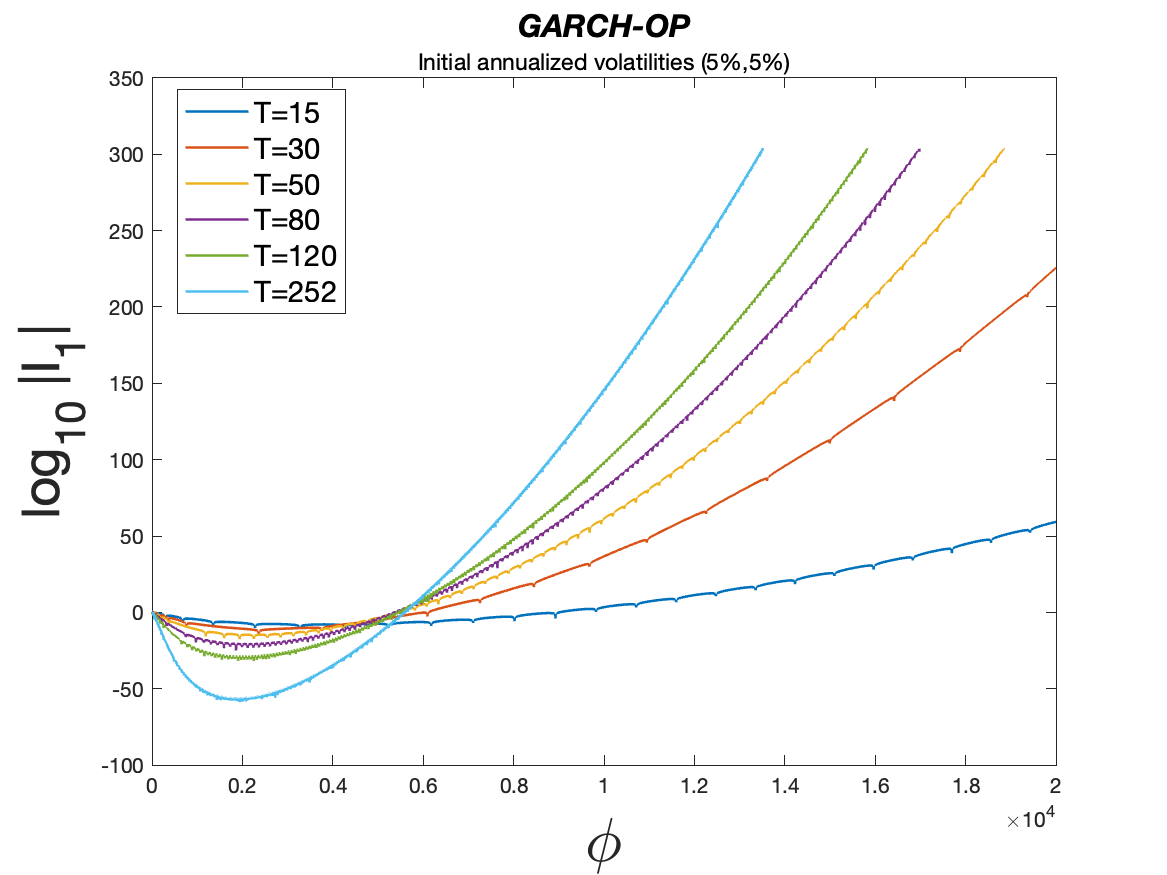}\hfill
        \includegraphics[width=0.49\textwidth]{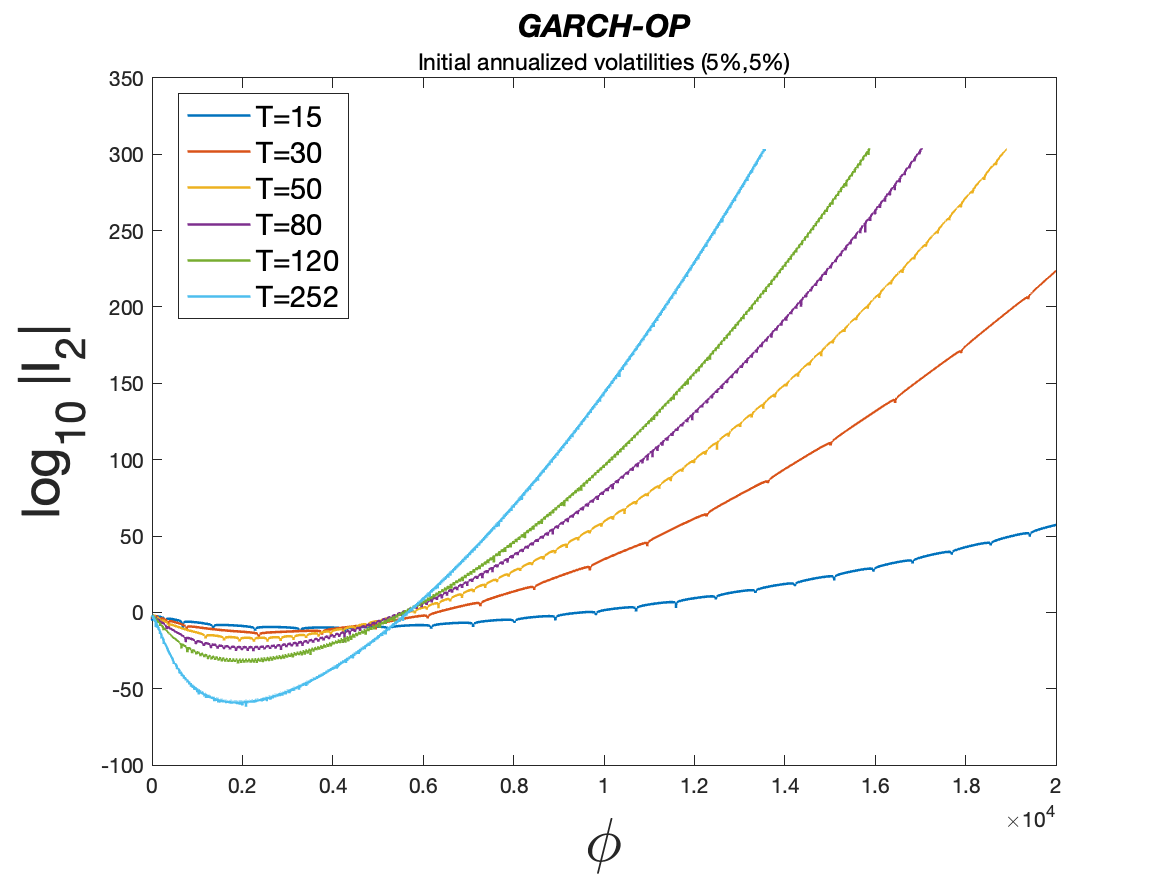}
        %\label{fig:plot1}
    \end{subfigure}

    \vspace{0.5cm}

    \begin{subfigure}{\textwidth}
        \centering
        \includegraphics[width=0.49\textwidth]{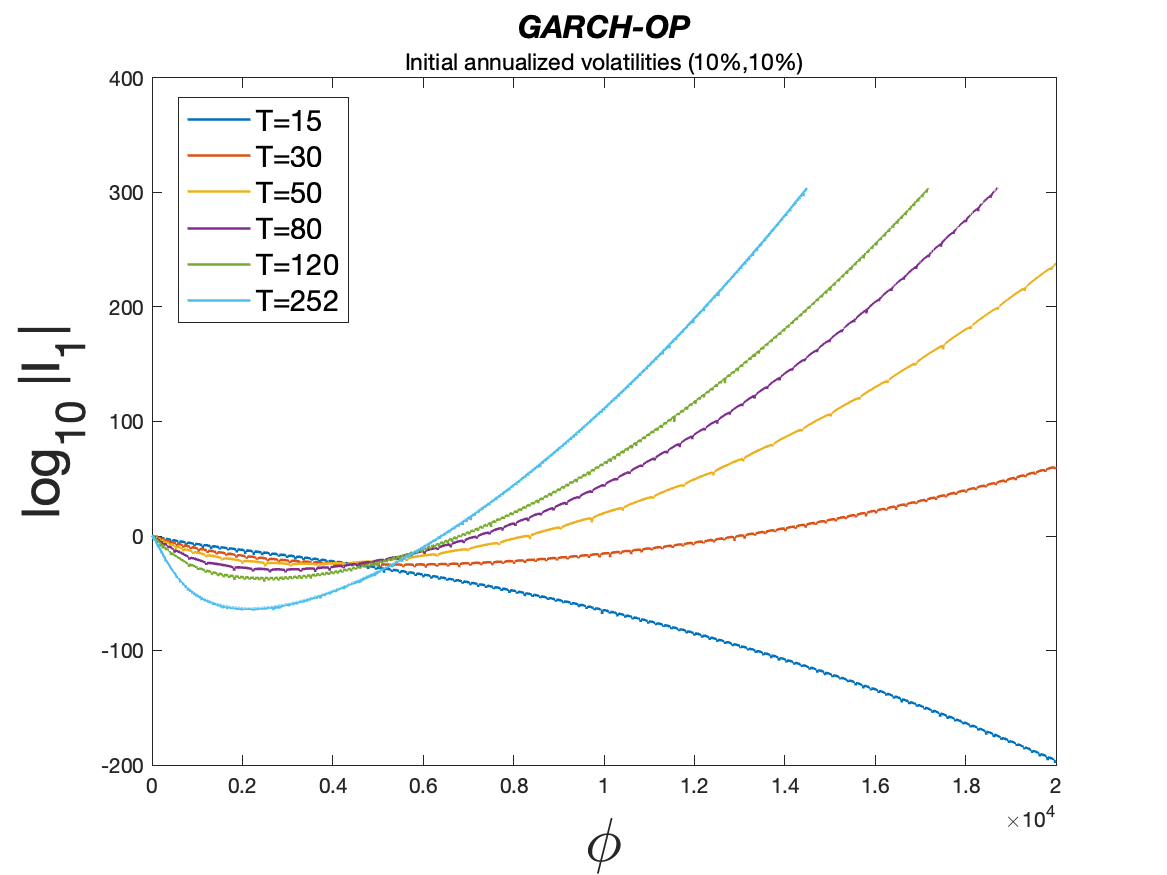}\hfill
        \includegraphics[width=0.49\textwidth]{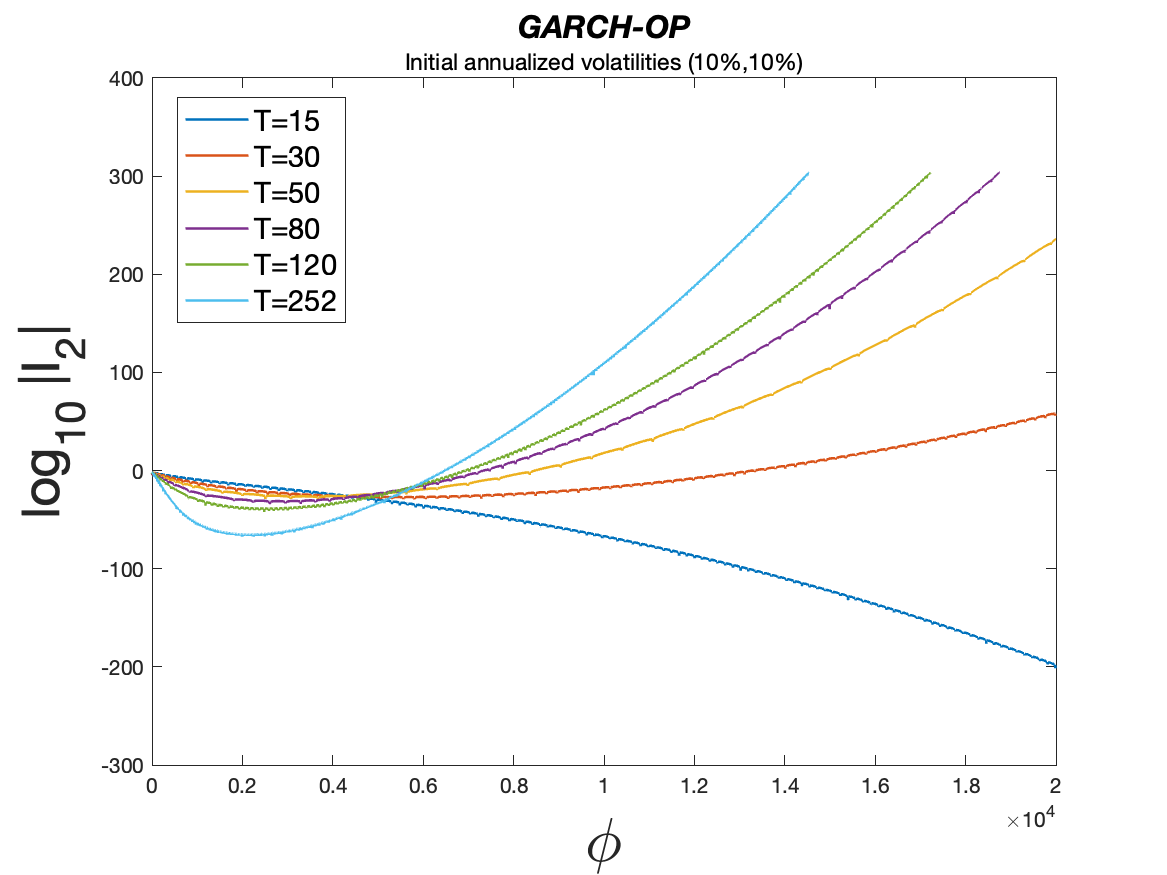}
        %\label{fig:plot3}
    \end{subfigure}
\end{figure}

Figures \ref{fig:diverg_hn_05}-\ref{fig:diverg_park} show the behavior of $|I_1|$ and $|I_2|$ for the 
\textit{GARCH-HN}, \textit{GARCH-CJOW} and \textit{GARCH-OP} models. In particular, Figures \ref{fig:diverg_hn_05} and \ref{fig:diverg_hn_10} display the behavior of $|I_1|$ and $|I_2|$, respectively, for the \textit{GARCH-HN} model using the parameters \textit{CJOW08} and \textit{CCLT23} in Table \ref{tab:param}. As the initial (annualized) volatility state we set $\sqrt{252 h_0}$ equal to either $5\%$ or $10\%$ and we set $r=10^{-5}$ and $S_0=K=100$. Instead, for the \textit{GARCH-CJOW} model we show the behavior of $|I_1|$ and $|I_2|$ in Figures \ref{fig:diverg_cjow_05} and \ref{fig:diverg_cjow_10} and for the \textit{GARCH-OP} model in Figure \ref{fig:diverg_park}. As for the initial volatility states, again we set $\sqrt{252 h_0}$ and $\sqrt{252 q_0}$ equal to $5\%$ and $10\%$, and we set $r=10^{-5}$ and $S_0=K=100$.

As we may notice, for both the \textit{CJOW08} and \textit{CCLT23} parameters, the \textit{GARCH-HN} model does not show any integrability issue for $I_1$ and $I_2$ in equation \eqref{eq_integrands}, whereas Figures \ref{fig:diverg_cjow_05}, \ref{fig:diverg_cjow_10} and \ref{fig:diverg_park} clearly show that $I_1$ and $I_2$ explode for almost all the considered maturities. Only for $T=15$ and for the $10\%$ initial volatility state both the \textit{GARCH-CJOW} and \textit{GARCH-OP} models do not exhibit any issues.

\section{An improved two-component GARCH model for option valuation} \label{sec_positive}

In this section, we propose a component volatility model that enhances the one developed by CJOW by ensuring that the volatility remains positive. We name this model \textit{GARCH-CPC}, which stands for Corrected Positive Component GARCH. The model consists of the following equations:

\begin{align}\label{eq_model_pos_rt}
    R_{t+1} &= r + \lambda h_{t+1} + \sqrt{h_{t+1}} Z_{t+1}, \\ \label{eq_model_pos_ht}
    h_{t+1} &= q_{t+1} + \Tilde{\beta} ( h_{t} - q_{t} ) + \alpha \left \{ \left( Z_{t}-\gamma_1 \sqrt{h_{t}} \right)^2 - \gamma^2_1 q_{t} \right\}  , \\ 
    q_{t+1} &= \omega + \rho q_{t} + \varphi \left(Z_{t}-\gamma_2 \sqrt{ h_{t} } \right)^2,
    \label{eq_model_pos_qt}
\end{align}

\noindent where $Z_t \overset{iid}{\sim} N(0,1)$ and $q_t$ represents a process related to the long-run volatility component as in the \textit{GARCH-CJOW} model. To ensure the positivity of $h_{t}$, we assume the following conditions for the parameters: $\omega \geq 0$, $\alpha >0$, $\varphi > 0$ and $ \Tilde{\beta} + \alpha \gamma^2_1 < \rho < 1$. Substituting equation \eqref{eq_model_pos_qt} into \eqref{eq_model_pos_ht} immediately shows that these parameter conditions are sufficient to guarantee the positivity of $h_t$ for all $t$. The long-run means are given by $\begin{pmatrix}
        \mathbb{E}[h_t] &
        \mathbb{E}[q_t]
    \end{pmatrix}^\top =  (I_2 - P)^{-1} R$, 
    where
    $I_2$ denote a $2\times2$ identity matrix, $R= \begin{pmatrix}
     \omega + \alpha + \varphi \\ 
     \omega + \varphi
 \end{pmatrix}$ and $P= \begin{pmatrix}
     \Tilde{\beta} + \alpha \gamma^2_1 + \varphi \gamma^2_2& \rho - \Tilde{\beta} - \alpha \gamma^2_1 \\
     \varphi \gamma^2_2 & \rho 
 \end{pmatrix}$. 
 
 We note that, for the model in equations \eqref{eq_cjow_ht}-\eqref{eq_cjow_qt}, $ \mathbb{E} [h_t - q_t] = 0$. In contrast, for 
 equations \eqref{eq_park_ht}-\eqref{eq_park_qt} it holds that $\mathbb{E} [h_t - q_t] = {\alpha}/(1- \Tilde{\beta} )$ and for equations \eqref{eq_model_pos_ht}-\eqref{eq_model_pos_qt} we have $\mathbb{E} [h_t - q_t] = {\alpha}/(1- \Tilde{\beta} - \alpha \gamma^2_1)$. Therefore, unlike in the \textit{GARCH-CJOW} model, the long-run volatility component in the \textit{GARCH-OP} and \textit{GARCH-CPC} models is represented not by \(q_t\), but rather by \(q_t + \Bar{h}\), where \(\Bar{h} = \frac{\alpha}{1 - \Tilde{\beta}}\) for the \textit{GARCH-OP} model, and \(\Bar{h} = \frac{\alpha}{1 - \Tilde{\beta} - \alpha \gamma^2_1}\) for the \textit{GARCH-CPC} model.

%We derive the unconditional mean of $h_t$ and $q_t$ as

%\begin{equation}
  %  \mathbb{E}[h_t] = \frac{  \left( \frac{ \omega + \varphi }{1-\rho} \right)  (1- \Tilde{\beta} -\alpha\gamma^2_1) + \alpha}{ (1 - \Tilde{\beta} - \alpha\gamma^2_1) ( 1 - \frac{\varphi \gamma^2_2}{1-\rho}) } \quad \text{and} \quad \mathbb{E}[q_{t}] = \frac{\omega  + \varphi  + \alpha \gamma^2_1 \mathbb{E}[h_t]}{1-\rho}. 
%\end{equation}

\subsection{The risk-neutral \textit{GARCH-CPC} dynamics}

Following \cite{christoffersen08}, we risk-neutralize the dynamics for the \textit{GARCH-CPC} model as follows:

\begin{align}\label{eq_model_pos_rt_rn}
    R_{t+1} &= r - \frac{1}{2} h_{t+1} + \sqrt{h_{t+1}} Z^*_{t+1}, \\ \label{eq_model_pos_ht_rn}
    h_{t+1} &= q_{t+1} + \Tilde{\beta} ( h_{t} - q_{t} ) + \alpha \left( Z^*_{t}-\gamma^*_1 \sqrt{h_{t}} \right)^2 - \alpha (\gamma^*_1)^2 q_{t}, \\ 
    q_{t+1} &= \omega + \rho q_{t} + \varphi \left(Z^*_{t}-\gamma^*_2 \sqrt{ h_{t} } \right)^2,
    \label{eq_model_pos_qt_rn}
\end{align}

\noindent with $\gamma^*_i = \gamma_i + \frac{1}{2} + \lambda $ for $i=1,2$ and $Z^*_{t} \sim N(0,1)$.

\subsection{Moment generating function}

To perform option valuation, we derive the moment generating function associated to the terminal log-spot price of equation \eqref{eq_model_pos_rt_rn}. 

\begin{prop} \label{prop1}
    The conditional moment generating function for the logarithm of the terminal log-price $\ln S_T$, denoted by $f(t,T;\phi)$, can be computed as 
\begin{equation} \label{eq_guess}
    f(t,T;\phi) = e^{\phi \log S_t + A_t(\phi) + B_{1,t}(\phi) (h_{t+1} - q_{t+1}) + B_{2,t}(\phi) q_{t+1}}, \quad t < T,
\end{equation}

\noindent where the expressions for $A_t(\phi)$, $B_{1,t}(\phi)$ and $B_{2,t}(\phi)$ are derived in equation \eqref{coeff_ricorsivi} in the Appendix.
\end{prop}

\section{Empirical results for returns and options} \label{sec_empirical}

In this section, we examine the goodness-of-fit of the \textit{GARCH-CJOW}, \textit{GARCH-OP} and the \textit{GARCH-CPC} models on both returns and options data.

The data for the daily levels of Adjusted Close Price for the S\&P500 returns series is retrieved from Refinitiv Datastream. As a proxy of the risk-free interest rate, we utilize the 3-month Treasury Bill rate and we gathered its time series data from the Federal Funds Effective Rate (FRED) dataset. For the returns, we considered two different periods: one covering the same period considered by CJOW, from July 2, 1962, to December 31, 2001 (labeled as ``Period 1") resulting in 9943 days, and another covering more recent data from January 2, 2002, to December 29, 2023 (labeled as ``Period 2") for a total of 5537 days.

We consider both Put and Call European options written on the S\&P500 index, with data retrieved from Thomson Reuters Eikon Datastream. The options we examined have maturities ranging from 2020 to 2023 and we consider the option daily prices from February 10, 2020, to December 29, 2023. As a common practice, following \cite{christoffersen12}, \cite{enzo23}, and \cite{enzo24}, we apply several exclusion filters retaining a total of 14,247 options prices. In particular, we keep only the options with time-to-maturity between 14 and 365 days and we select only out-of-the-money Put and Call options (we compute the moneyness as $K/S_t$, where $K$ is the strike price and $S_t$ is the underlying index level), and we filter out illiquid quotes by selecting only the six most liquid strikes at each maturity, and we consider option quotes only on Wednesday. Finally, we remove price quotes lower than $3.8\$$.

Following CJOW, the estimation we conducted relies on maximum likelihood estimation using only the log-returns. The results for Period 1 and Period 2 are shown in Table \ref{tab:tab_insample}. For Period 1, since we considered the same time frame as CJOW, we included the parameters they estimated in their article. To evaluate the goodness-of-fit on the returns data, we also include log-likelihood values alongside standard information criteria such as AIC and BIC. Finally, the fit to the option data is assessed using a standard metric, the implied volatility root mean square error:

\begin{equation}
    \text{IVRMSE(\%)} = 100 \times \sqrt{ \frac{1}{N} \sum_{i=1}^N \left( \textit{IV}^{MKT}_i - \textit{IV}^{MOD}_i \right)^2},
\end{equation}

\noindent where $\textit{IV}^{MKT}_i$ and $\textit{IV}^{MOD}_i$ denote the market and the model implied volatilities of the $i$-th option price, respectively.

Generally speaking, for both periods,  the \textit{GARCH-CPC} and \textit{GARCH-OP} models exhibit slightly worse fit on the returns data compared to the \textit{GARCH-CJOW}, as indicated by AIC and BIC values. However, for the \textit{GARCH-CJOW} model it was not possible to calculate all the option prices due to the issues of negative variance outlined previously in Section \ref{sec_opt_pricing}.

%In contrast to the \textit{GARCH-CJOW} model, not all estimated parameters of the \textit{GARCH-CPC} model are significant at the levels considered, indicating that the reaction to unpredictable shocks is not significantly different between the two components.

Thus, the fit to option data was assessable only for the \textit{GARCH-CPC} and \textit{GARCH-OP} models. In particular, the \textit{GARCH-CPC} displayed IVRMSE values of 5.7\% and 4.9\% for Period 1 and Period 2, respectively. For the \textit{GARCH-CJOW} model the ``NaN" values, reported at the bottom of Table \ref{tab:tab_insample}, indicate that it was not possible to compute prices for all the $N$ options based on the numerical integration of \eqref{eq_opt_price}. Whereas, for the parameters estimated in 
Table \ref{tab:tab_insample}, 
the \textit{GARCH-OP} did not encounter any issues in pricing the options we considered but returned higher pricing errors compared to the \textit{GARCH-CPC} model.

\begin{table}[h!]
\centering
\begin{threeparttable}
    \caption{Maximum likelihood estimation results}
    \label{tab:tab_insample}
    \begin{tabular}{@{} r 
                    S[table-format=-1.3e+2] 
                    S[table-format=-1.3e+2] 
                    S[table-format=-1.3e+2] 
                    S[table-format=-1.3e+2] 
                    S[table-format=-1.3e+2] 
                    S[table-format=-1.3e+2] @{}}
    \toprule
        & \multicolumn{3}{c}{{Period 1}} & \multicolumn{3}{c}{{Period 2}} \\
        \cmidrule(lr){2-4} \cmidrule(lr){5-7}
        & {\textit{GARCH-CPC}} & {\textit{GARCH-OP}} & {\textit{GARCH-CJOW}} & {\textit{GARCH-CPC}} & {\textit{GARCH-OP}} & {\textit{GARCH-CJOW}} \\
    \midrule
    $\omega$ 
        & {1.546e-16} & {8.678e-12} & {8.208e-07}\tnote{***} & {6.177e-14} & {6.689e-09} & {7.735e-07}\tnote{***} \\
        & {(7.420e-09)} & {(1.911e-08)} & {(7.620e-08)} & {(3.860e-09)} & {(5.651e-09)} & {(1.957e-07)} \\
    \addlinespace
    $\alpha$ 
        & {2.923e-06}\tnote{***} & {1.337e-06}\tnote{***} & {1.580e-06}\tnote{***} & {1.003e-06}\tnote{**} & {3.004e-06}\tnote{***} & {3.520e-06}\tnote{***} \\
        & {(1.392e-06)} & {(1.339e-07)} & {(2.430e-07)} & {(5.463e-07)} & {(6.094e-07)} & {(1.234e-06)} \\
    \addlinespace
    $\gamma_{1}$ 
        & {140.269}\tnote{***} & {438.588}\tnote{***} & {4.151e+02}\tnote{**} & {343.652}\tnote{***} & {337.450}\tnote{***} & {227.209}\tnote{***} \\
        & {(29.868)} & {(73.865)} & {(6.341e+01)} & {(123.759)} & {(59.820)} & {(82.479)} \\
    \addlinespace
    $\Tilde{\beta}$ 
        & {0.374}\tnote{***} & {0.776}\tnote{***} & {6.437e-01}\tnote{***} & {0.626}\tnote{***} & {0.887}\tnote{***} & {0.704}\tnote{***} \\
        & {(0.151)} & {(0.180)} & {(2.759e-02)} & {(0.232)} & {(0.033)} & {(0.082)} \\
    \addlinespace
    $\varphi$ 
        & {2.205e-06}\tnote{***} & {2.152e-06}\tnote{***} & {2.480e-06}\tnote{***} & {5.146e-06}\tnote{***} & {1.684e-06}\tnote{**} & {1.510e-06}\tnote{***} \\
        & {(4.226e-07)} & {(1.035e-07)} & {(1.160e-07)} & {(1.016e-06)} & {(8.926e-07)} & {(3.910e-07)} \\
    \addlinespace
    $\gamma_{2}$ 
        & {134.469}\tnote{***} & {58.924}\tnote{***} & {6.324e+01}\tnote{***} & {148.223}\tnote{***} & {120.697}\tnote{***} & {188.654}\tnote{***} \\
        & {(16.244)} & {(15.322)} & {(5.300)} & {(21.761)} & {(40.545)} & {(65.588)} \\
    \addlinespace
    $\rho$ 
        & {0.925}\tnote{***} & {0.960}\tnote{***} & {9.896e-01}\tnote{***} & {0.836}\tnote{***} & {0.949}\tnote{***} & {0.993}\tnote{***} \\
        & {(0.009)} & {(0.033)} & {(9.630e-01)} & {(0.033)} & {(0.016)} & {(0.002)} \\
    \addlinespace
    $\lambda$ 
        & {0.472} & {0.843} & {2.092}\tnote{***} & {-2.957} & {-3.957}\tnote{***} & {-3.412}\tnote{***} \\
        & {(0.813)} & {(0.852)} & {(7.729e-01)} & {(1.415)} & {(1.405)} & {(1.264)} \\
    \midrule
    Log-lik. & {33,978} & {33,979} & {34,102} & {17,993} & {17,997} & {18,065} \\
    AIC & {-67,940} & {-67,942} & {-68,188} & {-35,970} & {-35,978} & {-36,114} \\
    BIC & {-67,882} & {-67,884} & {-68,130} & {-35,917} & {-35,925} & {-36,061} \\
    IVRMSE(\%) & {5.787} & {6.237} & {NaN} & {4.965} & {5.163} & {NaN} \\
    \bottomrule
    \end{tabular}

    \begin{tablenotes}
        \item \textit{Note:} The standard errors, reported in parenthesis, are computed by inverting the negative Hessian matrix evaluated at the optimum parameter values. Statistical significance: \textsuperscript{*} at 0.1, \textsuperscript{**} at 0.05, and \textsuperscript{***} at 0.01 levels.
    \end{tablenotes}
\end{threeparttable}
\end{table}

\begin{figure}[h]
    \centering
    \includegraphics[width=0.5\textwidth]{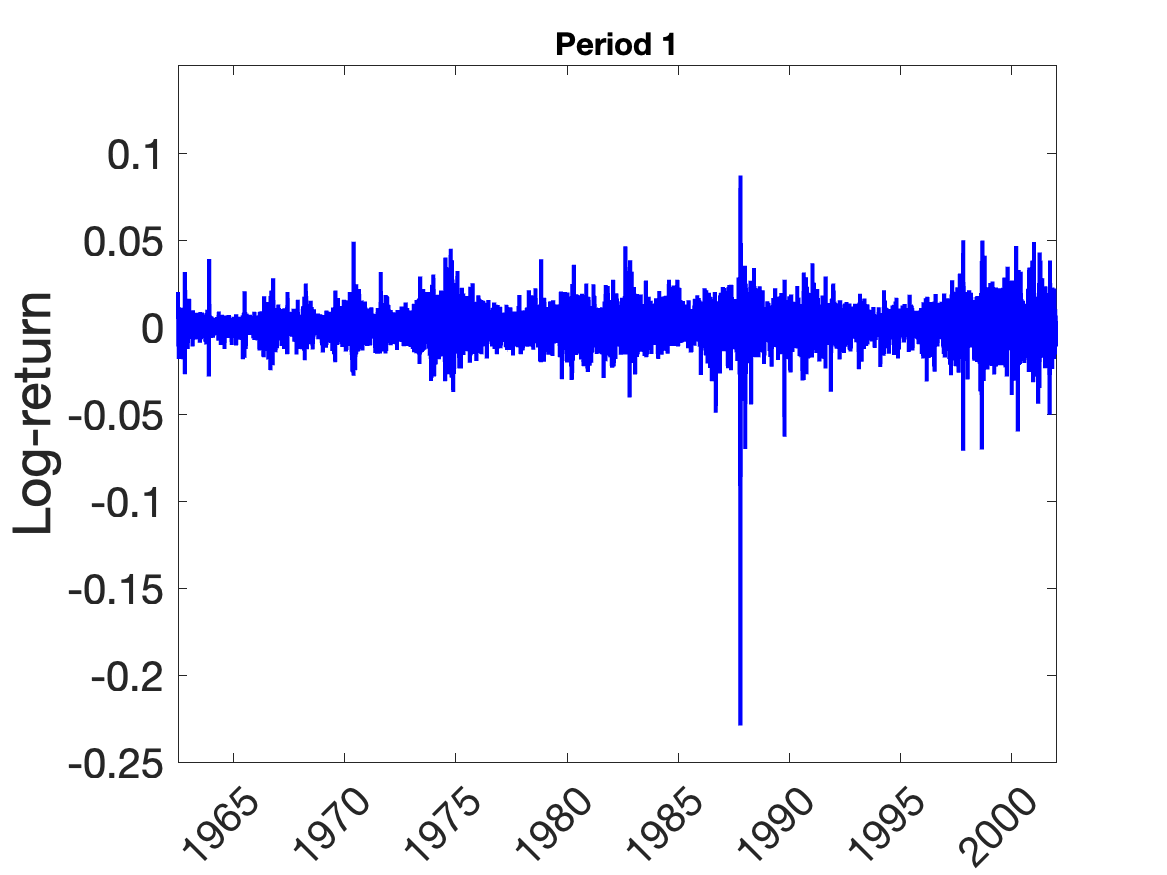}\hfill 
    \includegraphics[width=0.5\textwidth]{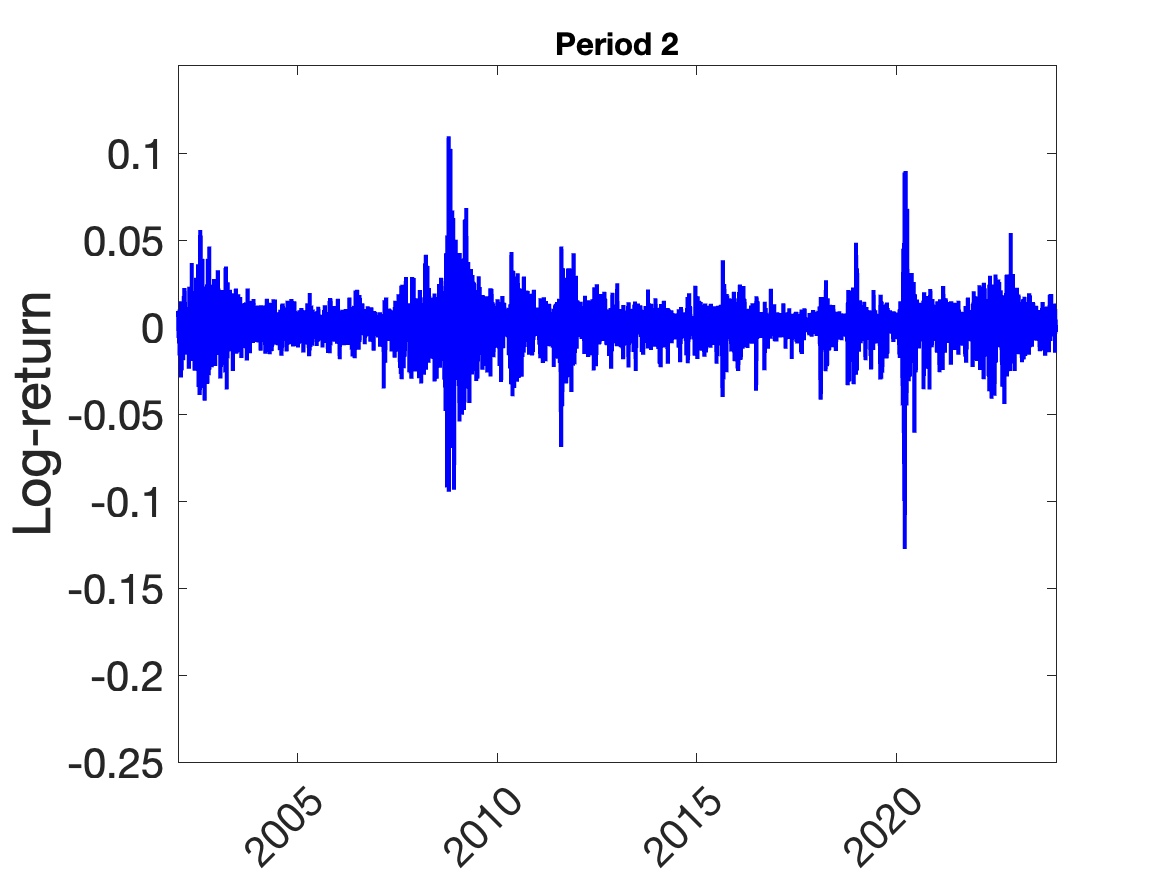}\hfill \\
    \includegraphics[width=0.5\textwidth]{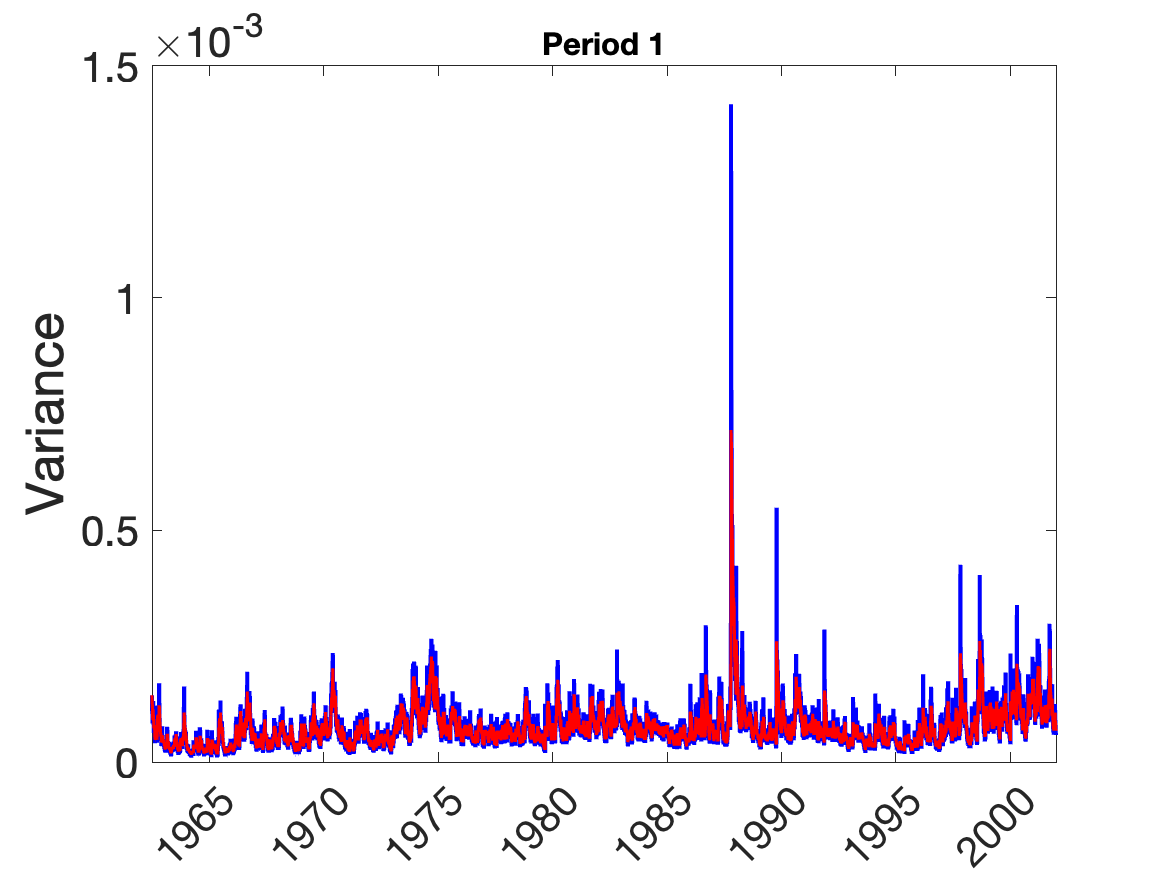}\hfill
    \includegraphics[width=0.5\textwidth]{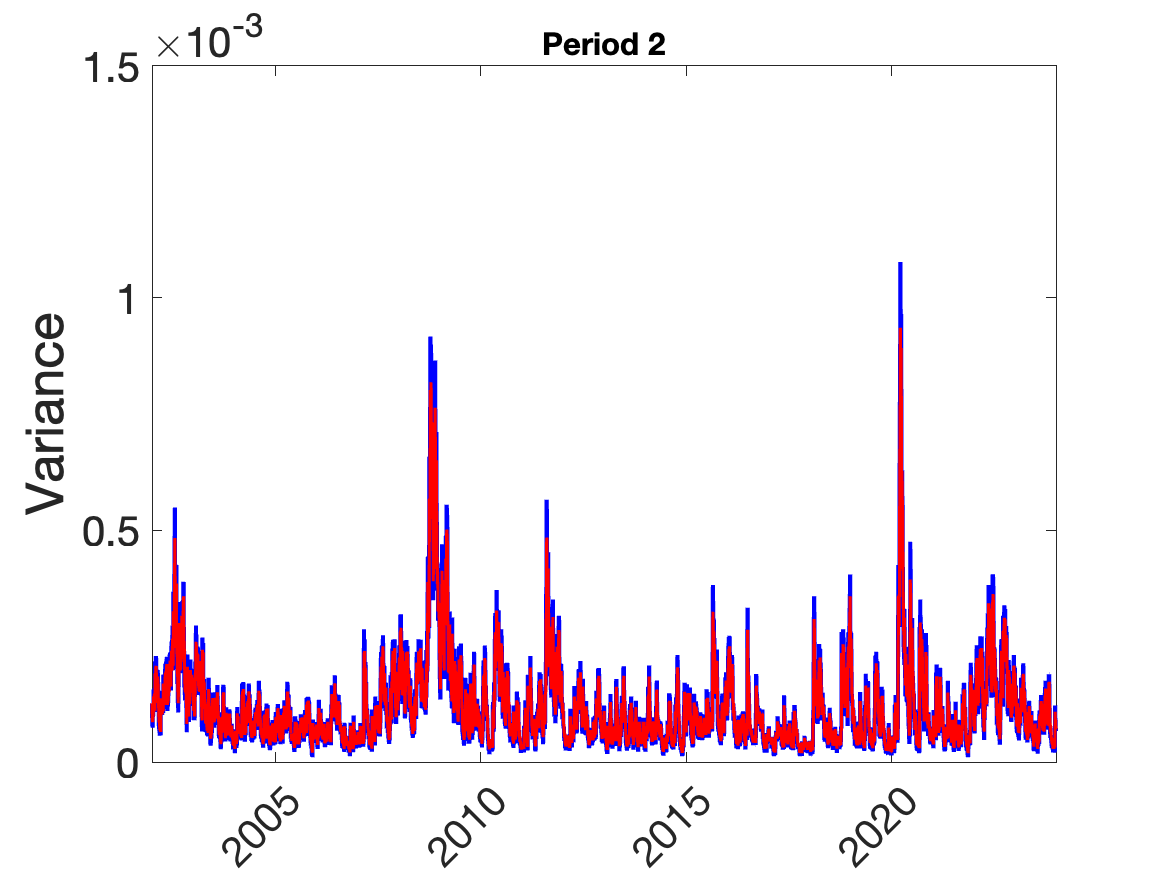}\hfill
    \caption{
    Log-returns S\&P 500 series (top panels) and \textit{GARCH-CPC} model (conditional) variance (bottom panels): $h_t$ (blue) and $q_t + \Bar{h}$ (red) where $\Bar{h} = {\alpha}/(1- \Tilde{\beta} - \alpha \gamma^2_1)$.}
    \label{fig:insample_periods}
\end{figure}

In Figure \ref{fig:insample_periods}, we display the S\&P500 daily log-return data for both Period 1 and Period 2 along with the filtered variance $h_t$ and the long-run component $q_t + \Bar{h}$ of the \textit{GARCH-CPC} model. Overall, the filtered long-run component is centered around the return variance $h_t$ and the model seems to correctly capture the spikes corresponding to higher volatility periods.

\begin{figure}[h] 
    \centering
    \includegraphics[width=0.5\textwidth]{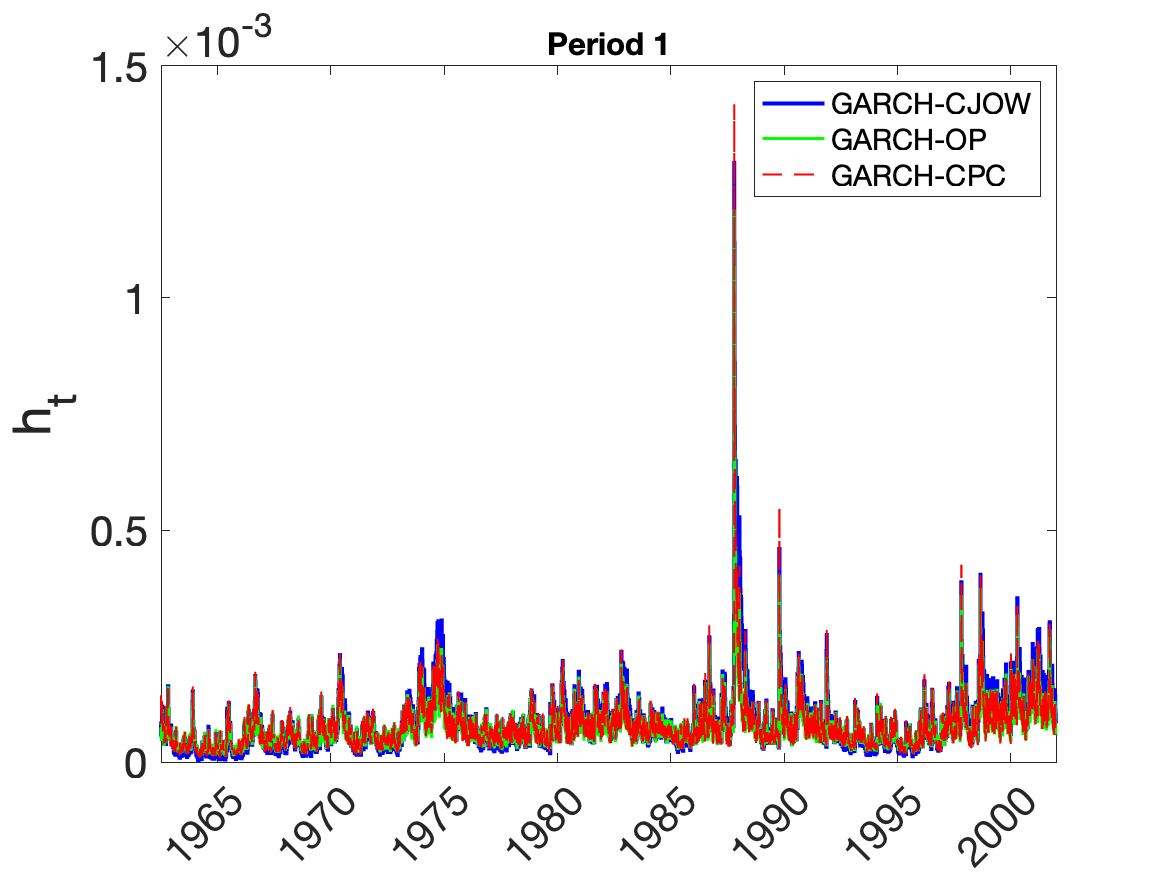}\hfill
    \includegraphics[width=0.5\textwidth]{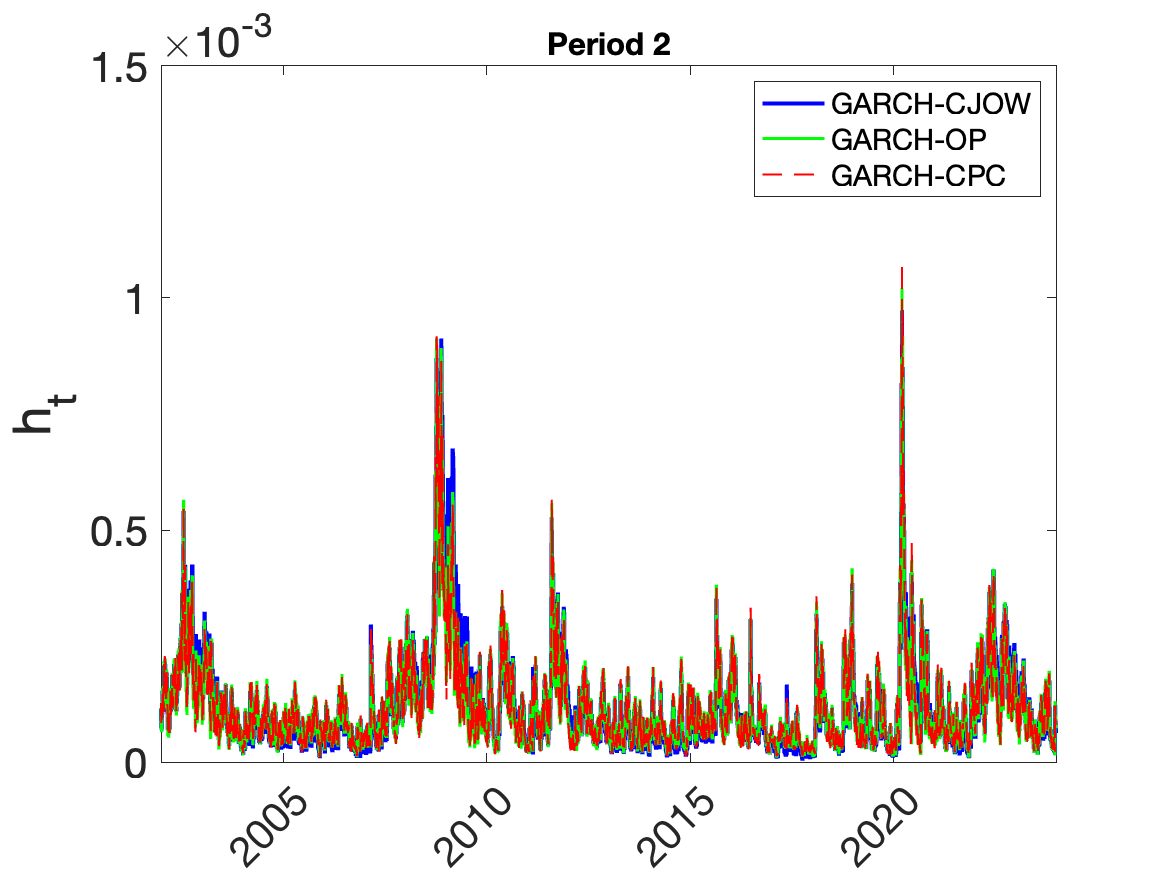}\hfill \\
   \includegraphics[width=0.5\textwidth]{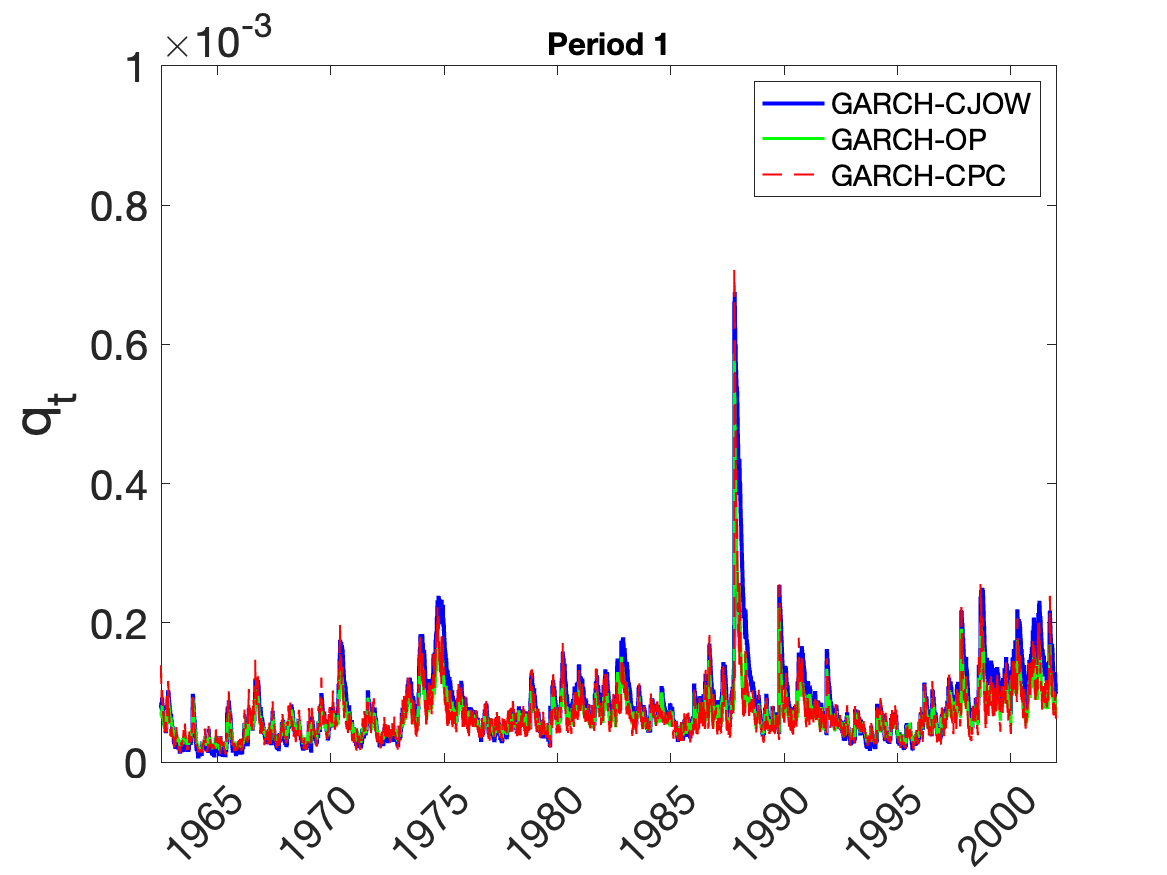}\hfill
\includegraphics[width=0.5\textwidth]{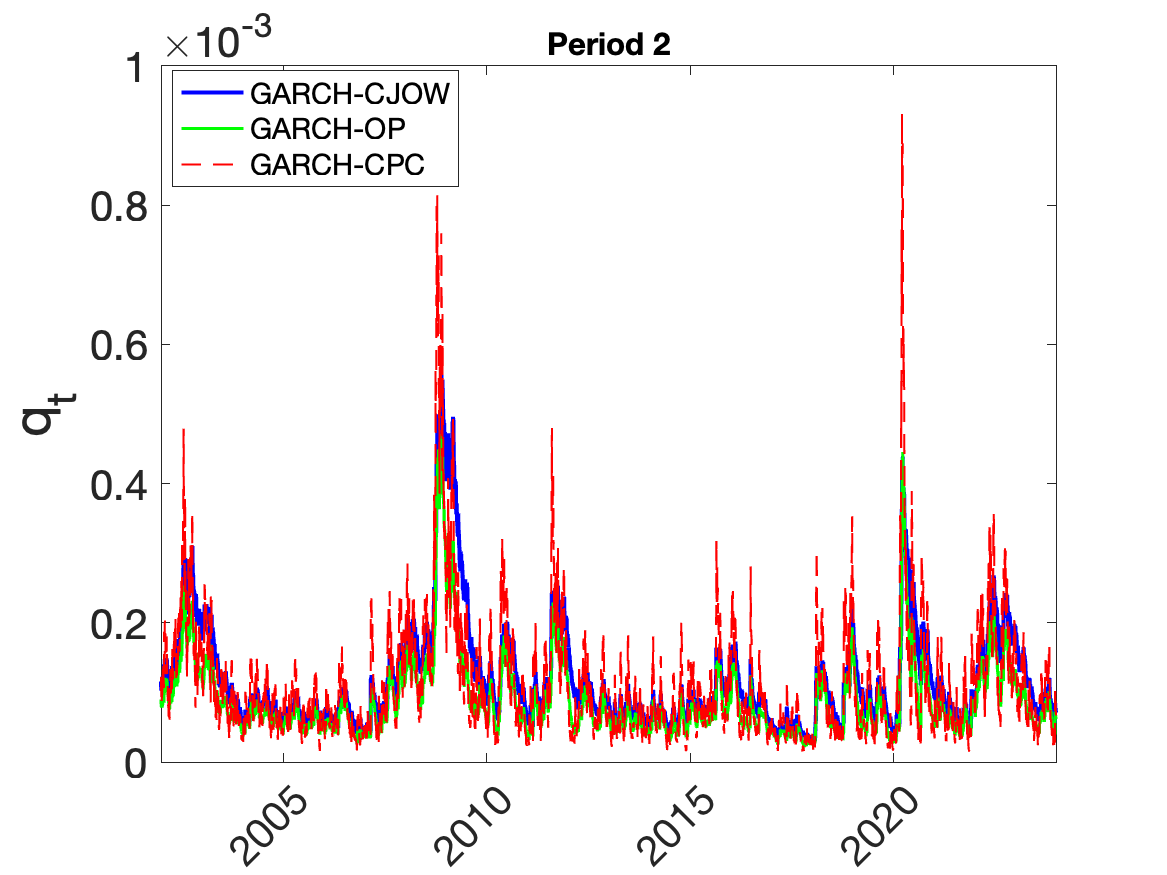}\hfill \\
\includegraphics[width=0.5\textwidth]{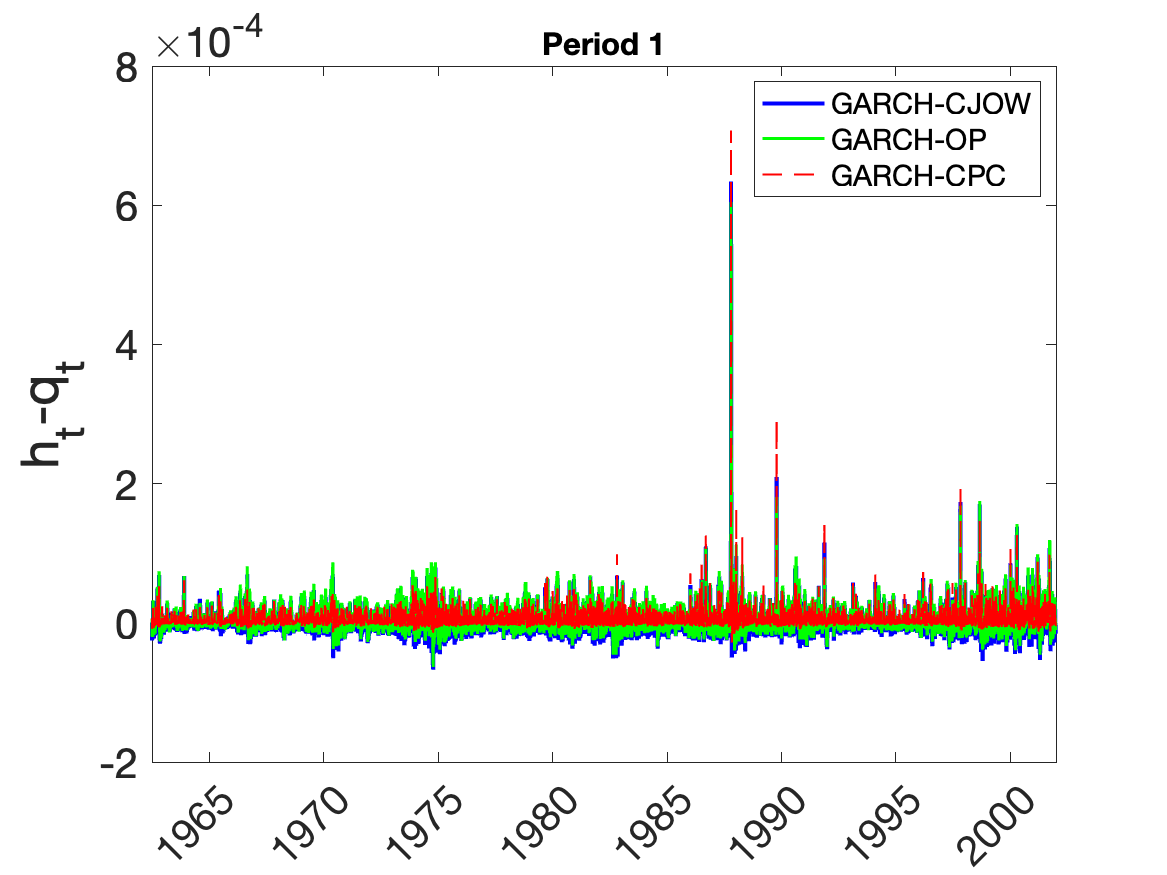}\hfill
\includegraphics[width=0.5\textwidth]{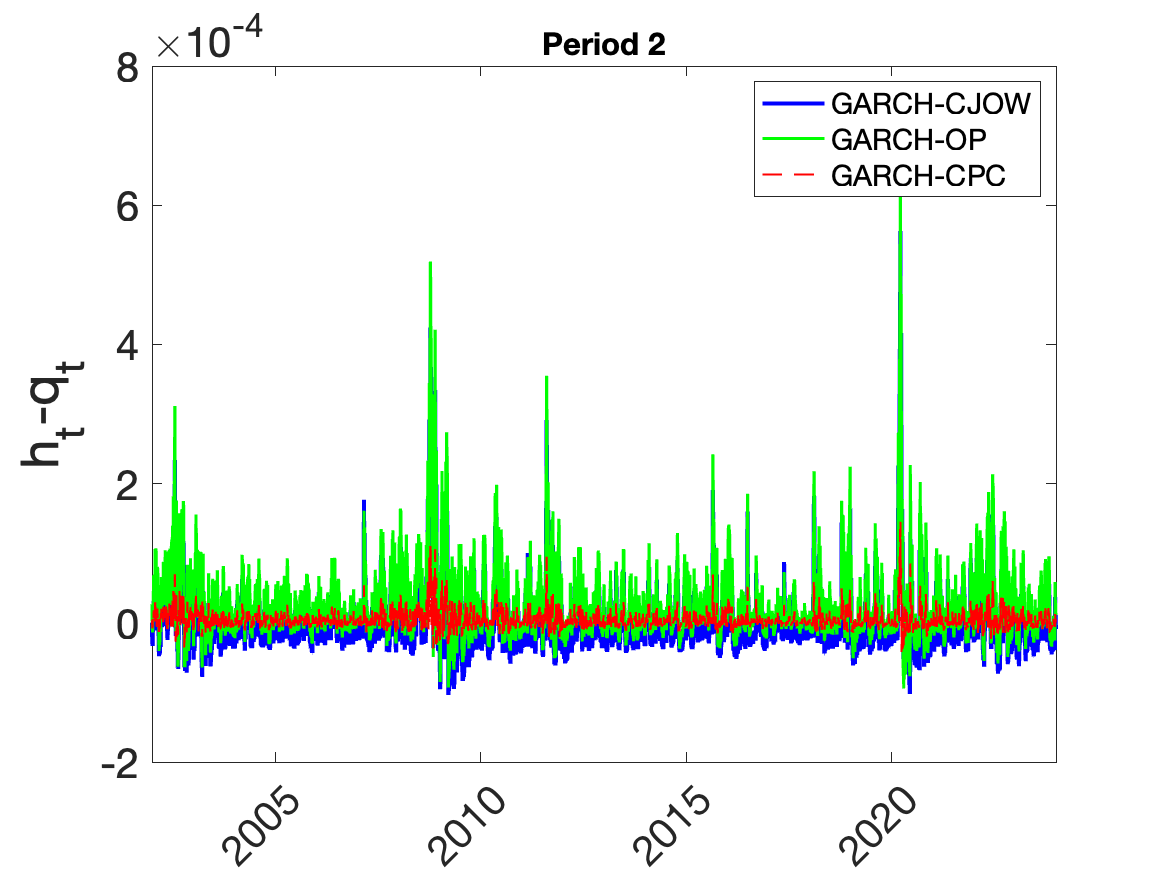}\hfill 
    \caption{Filtered  (conditional) variance.}
    \label{fig:variances}
\end{figure}

As a further comparison, in Figure \ref{fig:variances}, we display the filtered conditional variance $h_t$, the long-term component $q_t$, and the short-term component $h_t - q_t$, as considered by CJOW, for the \textit{GARCH-CJOW}, \textit{GARCH-OP}, and \textit{GARCH-CPC} models. The filtered variances are computed using the parameters from Table \ref{tab:tab_insample} for both Period 1 and Period 2. Overall, the filtered variance components of the three models exhibit similar patterns. Moreover, the short-term component $h_t - q_t$ for the \textit{GARCH-OP} and \textit{GARCH-CPC} models is not mean-zero, as explained in Section \ref{sec_positive}.

\section{Conclusions} 
\label{sec_conclusions}

In the influential article by \cite{christoffersen08}, it is demonstrated that the \textit{GARCH-CJOW} component model significantly outperforms single component models in capturing stock return volatility. Moreover, the \textit{GARCH-CJOW} model also shows excellent performance in explaining option prices.

However, a key limitation of the \textit{GARCH-CJOW} model is its failure to not guarantee positive volatilities, leading to substantial practical challenges. Specifically, even with realistic parameter sets, it is unfeasible to compute option prices for several maturities using the popular semi-closed formula, thereby vanishing one of the main advantages of affine models. 

These challenges were partially addressed by the model proposed in \cite{park23}. However, as we demonstrate in this paper, it still experiences negative volatilities. Therefore, we propose an enhancement of the \textit{GARCH-CJOW} and \textit{GARCH-OP} models, termed \textit{GARCH-CPC} model, which guarantees the positivity of the variance both theoretically and empirically. Our novel approach demonstrates comparable performance on returns data and superior performance in pricing options compared to both the \textit{GARCH-CJOW} and \textit{GARCH-OP} models.

\section*{Appendix}

\textit{Proof of Proposition \ref{prop1}}. To obtain the recursive equations for for $A_t (\phi)$, $B_{1,t}(\phi)$ and $B_{2,t}(\phi)$ for $t<T$, we use the tower property of the conditional expectation 

\begin{equation} \label{eq_tower}
f(t,T; \phi) = \mathbb{E} \left[  f(t+1,T;\phi) \middle| \mathcal{F}_t \right], 
\end{equation}

\noindent so that by using equation \eqref{eq_guess} we can express \eqref{eq_tower} as

\begin{equation} \label{eq_tower2}
    f(t,T;\phi) = \mathbb{E} \left[
    e^{\phi \log S_{t+1} + A_{t+1}(\phi) + B_{1,t+1}(\phi) \left( h_{t+2}-q_{t+2} \right)
    + B_{2,t+1}(\phi) q_{t+2} } \middle| \mathcal{F}_t \right]. 
\end{equation}

If we substitute equations \eqref{eq_model_pos_rt_rn}-\eqref{eq_model_pos_qt_rn} into \eqref{eq_tower2} we obtain

\begin{align*}
   f(t,T;\phi) = \mathbb{E} \biggl[
    &e^{\phi \left(\log S_t + r + \lambda h_{t+1} + \sqrt{ h_{t+1} } Z_{t+1}  \right)  +A_{t+1}(\phi)}   \\
    & \times e^{ B_{1,t+1}(\phi) \left( \Tilde{\beta} (h_{t+1} - q_{t+1}) + \alpha \left( Z^2_{t+1} + \gamma^2_1 h_{t+1}- 2 \gamma_{1} \sqrt{h_{t+1}} Z_{t+1} \right) - \alpha \gamma^2_1 q_{t+1}  \right) }  \\
    & \times e^{ B_{2,t+1}(\phi) \left( \omega + \rho q_{t+1} + \varphi \left( Z^2_{t+1} + \gamma^2_2 h_{t+1} - 2 \gamma_{2} \sqrt{h_{t+1}} Z_{t+1} \right) 
    \right) } 
    \biggl| \mathcal{F}_t \biggr]. 
\end{align*}

After re-arranging terms and performing some algebra, we obtain

\begin{align} \label{eq_cond_cf}
   f(t,T; \phi) = \mathbb{E} \biggl[
    &e^{\phi (\log S_t + r ) + \omega B_{2,t+1}(\phi)  + A_{t+1}(\phi) } \\ \notag
    & \times e^{\left(\phi \lambda  +  (\Tilde{\beta} + \alpha \gamma^2_1) B_{1,t+1}(\phi) + \varphi \gamma^2_2  B_{2,t+1}(\phi)  \right) (h_{t+1} - q_{t+1} ) + ( \phi \lambda + (\rho + \varphi \gamma^2_2 ) B_{2,t+1}(\phi) )   q_{t+1} } 
    \\ \notag
    & \times e^{  \left( \alpha B_{1,t+1}(\phi) + \varphi B_{2,t+1}(\phi) \right) Z^2_{t+1}
    - 2 \sqrt{h_{t+1}} Z_{t+1} \left( \alpha \gamma_1 B_{1,t+1}(\phi) + \varphi \gamma_2 B_{2,t+1}(\phi) - \frac{\phi}{2} \right) } 
    \biggl| \mathcal{F}_t \biggr], 
\end{align}

\noindent so that we can make use of the following result for a standard Normal random variable $Z$:
\begin{equation} \label{eq_mgf_trick}
    \mathbb{E}\left[ e^{ aZ^2 +bZ } \right]  = e^{\frac{b^2}{2(1-2a)} - \frac{1}{2} \log(1-2a)}, \quad a < \frac{1}{2}.
\end{equation}

Using \eqref{eq_mgf_trick} and by equating both sides of \eqref{eq_cond_cf} we obtain: 

\begin{align} \label{coeff_ricorsivi}
    A_t(\phi) &= A_{t+1}(\phi) + r \phi + B_{2,t+1}(\phi) \omega - \frac{1}{2} \log{(1-2(\alpha B_{1,t+1}(\phi) + \varphi B_{2,t+1}(\phi)))}, \\[5pt] \notag
    B_{1,t}(\phi) &= B_{1,t+1}(\phi) ( \Tilde{\beta} + \alpha \gamma^2_1) + \lambda \phi + B_{2,t+1}(\phi) \varphi \gamma^2_2 + 2 \frac{( \alpha \gamma_1 B_{1,t+1}(\phi) + \varphi \gamma_2 B_{2,t+1}(\phi) - \frac{\phi}{2})^2}{1-2(\alpha B_{1,t+1}(\phi) + \varphi B_{2,t+1}(\phi))}, \\[5pt] \notag
     B_{2,t}(\phi) &= B_{2,t+1}(\phi) (\rho + \varphi \gamma^2_2) + \lambda \phi +  2\frac{( \alpha \gamma_1 B_{1,t+1}(\phi) + \varphi \gamma_2 B_{2,t+1}(\phi) - \frac{\phi}{2})^2}{1-2(\alpha B_{1,t+1}(\phi) + \varphi B_{2,t+1}(\phi))}.
\end{align}

We can use equations \eqref{coeff_ricorsivi} to recursively calculate the coefficients starting from $A_T(\phi) = 0$, $B_{1,T}(\phi) = 0$ and  $B_{2,T}(\phi) = 0 $.
$\hfill \square$

\bibliographystyle{unsrtnat}
\bibliography{references}  %%% Uncomment this line and comment out the ``thebibliography'' section below to use the external .bib file (using bibtex) .

\begin{thebibliography}{11}
\providecommand{\natexlab}[1]{#1}
\providecommand{\url}[1]{\texttt{#1}}
\expandafter\ifx\csname urlstyle\endcsname\relax
  \providecommand{\doi}[1]{doi: #1}\else
  \providecommand{\doi}{doi: \begingroup \urlstyle{rm}\Url}\fi

\bibitem[Christoffersen et~al.(2008)Christoffersen, Jacobs, Ornthanalai, and Wang]{christoffersen08}
P.~Christoffersen, K.~Jacobs, C.~Ornthanalai, and Y.~Wang.
\newblock Option valuation with long-run and short-run volatility components.
\newblock \emph{Journal of Financial Economics}, 90\penalty0 (3):\penalty0 272--297, 2008.

\bibitem[Oh and Park(2023)]{park23}
D.H. Oh and Y.-H. Park.
\newblock {GARCH option pricing with volatility derivatives}.
\newblock \emph{Journal of Banking and Finance}, 146\penalty0 (2):\penalty0 106718, 2023.

\bibitem[Engle and Lee(1999)]{engle99}
R.F. Engle and G.~Lee.
\newblock {A long-run and short-run component model of stock return volatility. In: Engle, R.F. and White, H., Eds., Cointegration, Causality, and Forecasting: A Festschrift in Honor of Clive W.J. Granger}.
\newblock \emph{Oxford University Press}, pages 475--497, 1999.

\bibitem[Heston and Nandi(2000)]{heston00}
S.L. Heston and S.~Nandi.
\newblock {A closed-form GARCH option valuation model}.
\newblock \emph{The Review of Financial Studies}, 13\penalty0 (3):\penalty0 585--625, 2000.

\bibitem[Corsi et~al.(2013)Corsi, Fusari, and La~Vecchia]{corsi13}
F.~Corsi, N.~Fusari, and D.~La~Vecchia.
\newblock {Realizing smiles: Options pricing with realized volatility}.
\newblock \emph{Journal of Financial Economics}, 107\penalty0 (2):\penalty0 284--304, 2013.

\bibitem[Cheng et~al.(2023)Cheng, Chang, Lo, and Tsai]{cheng23}
H-W Cheng, L-H Chang, C-L Lo, and J.T. Tsai.
\newblock {Empirical performance of component GARCH models in pricing VIX term structure and VIX futures}.
\newblock \emph{Journal of Empirical Finance}, 72:\penalty0 122--142, 2023.

\bibitem[Bormetti et~al.(2015)Bormetti, Corsi, and Majewski]{bormetti15}
G.~Bormetti, F.~Corsi, and A.~Majewski.
\newblock {Smile from the past: A general option pricing framework with multiple volatility and leverage components}.
\newblock \emph{Journal of Econometrics}, 187\penalty0 (2):\penalty0 521--531, 2015.

\bibitem[Gil-Pelaez(1951)]{gilpelaez51}
J.~Gil-Pelaez.
\newblock {Note on the inversion theorem}.
\newblock \emph{Biometrika}, 38\penalty0 (3-4):\penalty0 481--482, 1951.

\bibitem[Christoffersen et~al.(2012)Christoffersen, Jacobs, and Ornthanalai]{christoffersen12}
P.~Christoffersen, K.~Jacobs, and C.~Ornthanalai.
\newblock {Dynamic jump intensities and risk premiums: Evidence from S\&P500 returns and options}.
\newblock \emph{Journal of Financial Economics}, 106\penalty0 (3):\penalty0 447--472, 2012.

\bibitem[Ballestra et~al.(2023)Ballestra, D'Innocenzo, and Guizzardi]{enzo23}
L.V. Ballestra, E.~D'Innocenzo, and A.~Guizzardi.
\newblock {Score-driven modeling with jumps: An application to S\&P500 returns and options}.
\newblock \emph{Journal of Financial Econometrics}, 22\penalty0 (2):\penalty0 375--406, 2023.

\bibitem[Ballestra et~al.(2024)Ballestra, D'Innocenzo, and Guizzardi]{enzo24}
L.V. Ballestra, E.~D'Innocenzo, and A.~Guizzardi.
\newblock A new bivariate approach for modeling the interaction between stock volatility and interest rate: An application to {S\&P}500 returns and options.
\newblock \emph{European Journal of Operational Research}, 314\penalty0 (3):\penalty0 1185--1194, 2024.

\end{thebibliography}





% \end{thebibliography}

\end{document}